\documentstyle [12pt,aps] {revtex}
\input epsf
\newcommand{\beq}{\begin{equation}}
\newcommand{\eeq}{\end{equation}}
\newcommand{\beqs}{\begin{eqnarray}}
\newcommand{\eeqs}{\end{eqnarray}}

\begin{document}

\baselineskip 6.0mm

\title{Spin dynamics simulations of the magnetic dynamics of RbMnF$_3$
and direct comparison with experiment}

\author {Shan-Ho Tsai\thanks{email: tsai@hal.physast.uga.edu}
\and Alex Bunker\thanks{current address: Max Planck Institute for Polymer Research,
Ackermann Weg 10, Mainz, Germany D-55021-3148, 
email: bunker@mpip-mainz.mpg.de}
\and D. P. Landau\thanks{email: dlandau@hal.physast.uga.edu}}

\address{
Center for Simulational Physics \\
The University of Georgia \\
Athens, GA 30602}

\maketitle

\begin{abstract}

\baselineskip 6.0mm

Spin-dynamics techniques have been used to perform large-scale simulations 
of the dynamic behavior of the classical Heisenberg antiferromagnet in
simple cubic lattices with linear sizes $L\leq 60$. This system is widely recognized
as an appropriate model for the magnetic properties of RbMnF$_3$. Time-evolutions 
of spin configurations were determined numerically from coupled equations of motion 
for individual spins using a new algorithm implemented by Krech {\it et al}, 
which is based on fourth-order Suzuki-Trotter decompositions of exponential operators. 
The dynamic structure factor was calculated from the space- and
time-displaced spin-spin correlation function. The crossover from hydrodynamic 
to critical behavior of the dispersion curve and spin-wave half-width was 
studied as the temperature was increased towards the critical temperature. The
dynamic critical exponent was estimated to be $z=(1.43\pm 0.03)$, which is 
slightly lower than the dynamic scaling prediction, but in good agreement with 
a recent experimental value. Direct, quantitative comparisons of both the 
dispersion curve and the lineshapes obtained from our simulations with very 
recent experimental results for RbMnF$_3$ are presented. 

\end{abstract}

\baselineskip 6.0mm

\pagestyle{empty}
\newpage

\pagestyle{plain}
\pagenumbering{arabic}
\renewcommand{\thefootnote}{\arabic{footnote}}
\setcounter{footnote}{0}

\section{Introduction}

Heisenberg models are  examples of magnetic systems that have true dynamics with the
real time dynamics governed by coupled equations of motion. According to the
classification of the different dynamic universality classes proposed by Hohenberg and 
Halperin in their work on the theory of dynamic critical phenomena  \cite{hohenhal},
the classical Heisenberg ferromagnet and antiferromagnet are of class J and G, respectively.
Although both classes have true dynamics, in the former class the order parameter 
(the uniform magnetization) is conserved,
whereas in the latter the order parameter (the staggered magnetization) is not a conserved
quantity. The difference in the dynamic behavior of the Heisenberg ferromagnet and antiferromagnet
can already be seen from the linear spin-wave theory, which predicts different low-temperature
dispersion curves for the two models. Dynamic critical
behavior is describable in terms of a dynamic critical exponent $z$ which depends on
conservation laws, lattice dimension, and the static critical exponents.  
The static critical behavior of three-dimensional Heisenberg models have been studied using a 
variety of approaches, including a high-resolution Monte Carlo simulation which determined 
the critical temperature and the static critical exponents for simple cubic and 
body-centered-cubic lattices \cite{kunTc}. In contrast, the theory of the dynamic behavior of 
Heisenberg models is not so well understood. 

A very close realization of an isotropic three-dimensional Heisenberg antiferromagnet is 
RbMnF$_3$. Early experimental studies \cite{pickart,teaney,windsor} have shown that 
in RbMnF$_3$ the Mn$^{2+}$ ions, with spin $S=5/2$, form a simple cubic lattice structure 
with a nearest-neighbor exchange constant of $J^{exp}=(0.58\pm 0.06)$ meV and a second-neighbor 
constant of less than $0.04$ meV [both defined using our convention for the exchange constant 
to be shown
in Eq. (\ref{hamq})]. Magnetic ordering with antiferromagnetic alignment of spins occurs below 
83K, which we denote as the critical temperature $T_c$. In this material, the magnetic 
anisotropy was found to be only about $6\times 10^{-6}$ of the exchange field, and no deviation 
from cubic symmetry was seen at $T_c$ \cite{teaney62,teaney63}. Both the static properties and 
the dynamic response of RbMnF$_3$ have been examined through neutron
scattering experiments. The early work of Tucciarone {\it et al} \cite{tucciarone} found that
in the critical region the neutron scattering function has a central peak (peak at zero
frequency transfer) and a spin-wave peak. Later, the experimental study by Cox {\it et al} 
\cite{cox} observed a small central peak below $T_c$ as well. The more recent experiments by 
Coldea {\it et al} \cite{coldea} have also found central peaks for $T\le T_c$, in 
agreement with previous experiments. From the theoretical side, renormalization-group (RNG) 
below $T_c$ \cite{mazenkoTb} predicts spin-wave peaks, and a central peak in the
longitudinal component of the neutron scattering function. However, at the critical temperature
both renormalization-group \cite{mazenkoTc} and mode-coupling \cite{cuccoli} theories
predict only the spin-wave peak, i.e. the central peak has not been predicted by these 
theories. The central peak is thought to be due
to spin diffusion resulting from nonlinearities in the dynamical equations \cite{tucciarone}. 
In their recent experiment, Coldea {\it et al} \cite{coldea} obtained the most
precise experimental estimate of the dynamic critical exponent, $z=(1.43\pm 0.04)$. 
The theoretical prediction \cite{hohenhal,mazenkoTc,cuccoli} is $z=1.5$ for class G
models in three dimensions. 

Large-scale computer simulations using spin-dynamics techniques to study the dynamic behavior
of Heisenberg ferromagnet and antiferromagnet have been carried out by Chen and Landau \cite{kun}
and Bunker {\it et al} \cite{alex}, respectively. So far, however, there are no direct comparisons 
of the dispersion curve and the dynamic structure lineshapes obtained from simulations with the 
corresponding experimental results.
In the present work we have carried out large-scale simulations of the dynamic
behavior of the Heisenberg antiferromagnet on a simple cubic lattice, and make direct comparisons
with experimental data for RbMnF$_3$. 
Sec. II of this paper contains the definition of the model and introduces some simulation 
background. In Sec. III we
present and discuss our simulation results and compare them with experiments.
Sec. IV contains some concluding remarks. 

\section{Model and Methods}

\subsection{Model}

The classical Heisenberg antiferromagnetic model is defined by the
Hamiltonian
\beq
{\cal H} = J \sum_{<{\bf rr'}>} {\bf S_r}\cdot {\bf S_{r'}},
\label{ham}
\eeq
where ${\bf S_r}=({S_{\bf r}}^x,{S_{\bf r}}^y,{S_{\bf r}}^z)$ is a three-dimensional
classical spin of unit length at site ${\bf r}$ and $J>0$ is the 
antiferromagnetic coupling constant between nearest-neighbor pairs
of spins. We consider $L\times L\times L$ simple cubic lattices with periodic
boundary conditions. The time dependence of each spin can be determined from the 
integration of the equations of motion [Eq.(2) in Ref.\cite{alex}], and 
the dynamic structure factor $S({\bf q},\omega)$ for momentum transfer
${\bf q}$ and frequency transfer $\omega$, observable in neutron scattering
experiments, is given by
\beq
S^k({\bf q},\omega)=\sum_{{\bf r,r'}} \exp[i {\bf q}\cdot ({\bf r}-{\bf r'})]
\int_{-\infty}^{+\infty} \exp(i\omega t) C^k({\bf r} - {\bf r'},t) \frac{dt}{\sqrt{2\pi}},
\eeq
where $C^k({\bf r} - {\bf r'},t)$ is the space-displaced, time-displaced spin-spin correlation
function defined, with $k=x, y,$ or $z$, as
\beq
C^k({\bf r} - {\bf r'},t) =\langle {S_{{\bf r}}}^k(t){S_{{\bf r'}}}^k(0)\rangle-
\langle {S_{{\bf r}}}^k(t)\rangle\langle {S_{{\bf r'}}}^k(0)\rangle.
\eeq
The displacement ${\bf r}$ is in units of the lattice unit cell length $a$. 
In the case of antiferromagnets, the wave-vectors are measured with respect to the 
$(\pi,\pi,\pi)$ point which corresponds to the Brillouin zone center.

\subsection{Simulation method}

Using a combination of Monte Carlo and spin-dynamics methods\cite{kun,alex,gerling}, we simulated
the behavior of the simple-cubic classical Heisenberg antiferromagnet with $12\le L\le 60$
at the critical temperature $T_c=1.442929(77)J$ \cite{kunTc} and below $T_c$. We have chosen 
units such that the Boltzmann constant $k_B=1$. 

Equilibrium configurations were generated using a hybrid Monte Carlo method 
in which a single hybrid Monte Carlo step consisted of two Metropolis steps and eight 
overrelaxation steps \cite{kun,alex}. Typically
1000 hybrid Monte Carlo steps were used to generate each equilibrium configuration and
the coupled equations of motion were then integrated numerically, using these states 
as initial spin configurations.
Numerical integrations were performed to a maximum time $t_{max}$, using a
time step of $\Delta t$. The space-displaced, time-displaced
spin-spin correlation function $C^k({\bf r} - {\bf r'},t)$ was computed for time-displacements
ranging from $0$ to $t_{cutoff}$. Each of such correlations
was calculated from an average over between 40 to 80 different time starting points, evenly 
spaced by $10\Delta t$. 
Our unit for time is the interaction constant $J$ defined in Eq. (\ref{ham}). 
The parameters used in this work were as follows. At $T=0.9 T_c$ we used lattices sizes
$L=12,24,36,48$, and $60$, for which the respective time parameters were $(t_{max} J,
t_{cutoff} J) = (440,400), (440,400), (480,400), (680,600)$, and $(1080,1000)$. 
The respective numbers of initial configurations used were $7020$, $2850$, $1110$, $400$, and
$810$. At $T_c$ and lattice sizes 
$L=18,24,30,36,48$ we used $t_{max}J=480$ and $t_{cutoff}J=400$, and for $L=60$ we used
$t_{max}J=1080$ and $t_{cutoff}J=1000$. The number of initial configurations used at $T_c$ was
$1000$ for $L=18,30,36$, whereas for $L=24,48$, and $60$ the respective numbers were $1125$, 
$715$, and $510$.
Other temperatures considered, chosen to coincide with experimental values, were 
$T=0.774T_c$, $0.846T_c$ and $0.936T_c$, for which
$t_{max}J=480$, $t_{cutoff}J=400$ and the number of initial configurations used was $120$. 
For lattice size $L=24$ at $T=0.9T_c$ 
the integration was carried out with a time step 
$\Delta t=0.01J^{-1}$ using a fourth-order predictor-corrector method, as in Refs.\cite{kun,alex}.
For other lattice sizes and temperatures, we used a new algorithm\cite{krech} based on
fourth-order Suzuki-Trotter decompositions of exponential operators, with a time step 
$\Delta t=0.2J^{-1}$ (except for $L=12$ at $T=0.9T_c$ for which $\Delta t=0.1J^{-1}$).
The latter method allowed
us to use larger integration time steps and thus to carry out the integration to larger
$t_{max}$, as compared with previous work \cite{kun,alex}. In the original study of Chen
and Landau \cite{kun} $t_{max}=120J^{-1}$ and in the work of Bunker {\it et al}  
$t_{max}=200J^{-1}$. In comparison, in the present work we have used a larger 
$t_{max}\geq 440J^{-1}$, more equilibrium initial spin configurations,
and a larger lattice size, namely $L=60$. As it is clear from the above, we have 
concentrated 
our efforts on two cases: one temperature below $T_c$, namely $T=0.9 T_c$, and at $T_c$. 

We also used the technique of calculating partial spin sums ``on the
fly''\cite{kun,alex} which limits us to data in the $(q,0,0)$, $(q,q,0)$ and $(q,q,q)$ directions
with $q$ determined by the periodic boundary conditions,
\beq
q=\frac{2\pi n}{L},\qquad n=\pm 1,\pm 2,...,\pm (L-1),L.
\label{qpbc}
\eeq
Since all three Cartesian spatial directions are equivalent by symmetry, the same operation
carried out for the $(q,0,0)$ direction was also carried out for the other two reciprocal
lattice directions $(0,q,0)$ and $(0,0,q)$ and the results were averaged. Similarly, the same
operations carried out for the $(q,q,0)$ and $(q,q,q)$ directions were also carried
out for the equivalent reciprocal lattice directions and the results were 
averaged for each case. 

For the ferromagnetic case the total magnetization is conserved and the dynamic structure 
factor $S({\bf q},\omega)$ can be separated into a component along the axis of the
total magnetization (longitudinal component) and a transverse component. However, for 
the antiferromagnet the order parameter is not conserved and separation is not possible.
We thus refer to the average
\beq
S({\bf q},\omega)=\frac{1}{3}[S^x({\bf q},\omega)+S^y({\bf q},\omega)+S^z({\bf q},\omega)]
\label{sqw}
\eeq
as the dynamic structure factor. 

\subsection{Dynamic finite-size scaling}

Two practical limitations on spin-dynamics techniques imposed by limited computer resources
are the finite lattice size and the finite evolution time. The finite time cutoff can introduce
oscillations in $S({\bf q},\omega)$, which can be smoothed out by convoluting the spin-spin
correlation function with a resolution function in frequency. 
The finite-size scaling \cite{kun,rapaport} can be used to extract the dynamic critical 
exponent $z$ using
\beq
\frac{\omega\bar {S_L}^k({\bf q},\omega)}{{\bar{\chi}_L}^k({\bf q})}=
G(\omega L^z,qL,\delta_{\omega}L^z)
\eeq
and 
\beq
\bar\omega_m=L^{-z}\bar\Omega(qL,\delta_{\omega}L^z),
\label{omegam}
\eeq
where $\bar {S_L}^k({\bf q},\omega)$ is the dynamic structure factor convoluted with a Gaussian
resolution function with characteristic parameter $\delta_{\omega}$, 
$\bar\omega_m$ is a characteristic frequency, defined as
\beq
\int_{-\bar\omega_m}^{\bar\omega_m}\bar {S_L}^k({\bf q},\omega)\frac{d\omega}{2\pi}
=\frac{1}{2}{\bar{\chi}_L}^k({\bf q}),
\eeq
and ${\bar{\chi}_L}^k({\bf q})$ is the total integrated intensity.

For the time cutoffs of at least $400J^{-1}$ used in the present simulations, the oscillations 
in the dynamic structure factor due to the finite $t_{cutoff}$ were not very significant.
Thus, we first estimate the dynamic critical exponent $z$ without using a resolution function,
or equivalently, we take $\delta_{\omega}=0$. In this case the analysis is simpler and a value
for $z$ can be obtained from the slope of a graph of $\log(\omega_m)$ vs $\log(L)$ (where
$\omega_m$ is the characteristic frequency in the absence of a resolution function) if the
value $qL$ is fixed. It is important to note that lattice sizes included in this calculation
should be large enough to be in the asymptotic-size regime. 
The approximate lattice size of the onset of the asymptotic regime can be estimated by looking
at the behavior of $\omega L^z$ for different lattice sizes using trial values of $z$. 

The effects of the small oscillations in the dynamic structure factor 
on the dynamic critical exponent $z$ can be evaluated by repeating the analysis using a
resolution function. For this purpose, we chose
\beq
\delta_{\omega}=0.02{\left [ \frac{60}{L} \right ] }^z
\label{deltaom}
\eeq
so that the function $\bar\Omega(qL,\delta_{\omega}L^z)$ in Eq. (\ref{omegam}) is a constant
if $qL$ is fixed, yielding
\beq
\bar\omega_m\sim L^{-z}.
\label{omegamLz}
\eeq
Because $\delta_{\omega}$ depends on $z$, this exponent had to be determined iteratively.
We used $n=1,2$ and several initial values for $z$ in the iterations, in order to check how 
stable the converged value of $z$ is.

\section{Results\label{results}}
\subsection{Numerical data for $S({\bf q},\omega)$}

For $T\le T_c$ our results for the dynamic structure factor, as 
defined in Eq. (\ref{sqw}), show a spin-wave and a central peak. 
In Fig. \ref{lshape09} we show lineshapes for lattice size $L=60$ at 
$T=0.9T_c$ and wave-vectors $q=\pi/10$, $\pi/6$ and $\pi/3$ in the [100]  
direction. We see that as $q$ increases, the central peak broadens and its
relative amplitude increases.
Fig. \ref{lshapeTc} shows lineshapes for $L=60$ at $T_c$ and 
wave-vectors $q=\pi/10$ and $\pi/6$ in the [100] direction.
It is clear from these lineshapes that the oscillations due to the finite $t_{cutoff}$ are
indeed negligible; therefore, in our analysis of the lineshapes we have not convoluted our 
results with a resolution function. (As explained in the following section, we later 
convoluted our results with a Gaussian resolution function in order to directly compare our
lineshapes with the experiments. The reason for this is that there is an intrinsic finite 
resolution 
in the experimental data due to the finite divergence of neutron beams.) The width of any
structure in the lineshapes discussed here is much larger than our resolution in
frequency $\Delta\omega=1.2\pi/t_{cutoff}$.

Below $T_c$, previous theoretical \cite{mazenkoTb} and experimental \cite{coldea} 
studies motivated us to extract the position and the half-width of the spin-wave 
and central peaks by fitting the lineshape to a Lorentzian form
\beq
S({\bf q},\omega)=\frac{A\Gamma_1^2}{\Gamma_1^2+\omega^2}+\frac{B\Gamma_2^2}
{\Gamma_2^2+(\omega+
\omega_s)^2}+\frac{B\Gamma_2^2}{\Gamma_2^2+(\omega-\omega_s)^2}
\label{lorentz}
\eeq
where the first term corresponds to the central peak and the last two terms are contributions
from the spin-wave creation and annihilation peaks at $\omega=\pm\omega_s$. 
For $T=0.9T_c$ we find that 
Lorentzian lineshapes fit our results reasonably well for small values of $q$, except for 
the smallest value, namely $q=2\pi/L$, in the [100] direction. The reason is that for each 
lattice size $L$, the dynamic structure factor for the smallest 
value of $q$ corresponds to correlations between spins displaced in space by a distance $L/2$, 
and the effect of the finite lattice size is particularly prominent in these cases, causing
the lineshapes to depart significantly from a Lorentzian form. For large values of $q$
(approximately $q>2\pi(L/4-2)/L$) the Lorentzian form given in Eq. (\ref{lorentz}) does not
fit the data, especially at large frequency transfer. In general, the fitted parameters
varied when different frequency ranges were used in the fit. Although this variation was
small, it was often larger than the statistical error
in the fitted parameters obtained from the fit using a single frequency range.
Therefore, for $T=0.9T_c$ we estimated the error in the
fitted parameters by fitting the lineshapes using three different ranges of frequency. The values
of the parameters were then averaged and the error bars estimated from the fluctuations. 
At $T_c$, renormalization-group theory (RNG) \cite{mazenkoTc} predicts a non-Lorentzian functional 
form for the spin-wave lineshape, which has been used along with a Lorentzian central peak
to analyze experimental data \cite{coldea}.
Since it is more complicated to perform fits to this RNG functional form and since the 
spin-wave peaks obtained from the simulations are more pronounced than in the experiment,
and thus less dependent on the fitted functional, we have fitted the lineshapes at $T_c$ to 
Lorentzians, as given in Eq. (\ref{lorentz}). Although 
obtaining a good fit to our data at $T_c$  was more difficult than below $T_c$, the resulting fits
at $T_c$ are still reasonable. Unlike for $T=0.9T_c$, at $T_c$ the lineshape parameters used in the
analysis below are the values obtained from the fit to only one frequency range, which was the
one that gave the best fit. 
One should then expect that the actual error in the fitted parameters at $T_c$ is larger (by up
to a factor of 5) than the error bars shown in the figures below. 

In addition to the spin-wave and the central peaks, we observed some other peaks on the
large frequency tail of the spin-wave peaks. Although these large frequency peaks had very
small amplitudes, they could be discerned from the background fluctuations. Using the
spin-wave frequencies in the [100], [110] and [111] directions we could check that the 
position of these extra peaks corresponded to frequencies of two spin-wave addition peaks. 
These extra structures in the lineshapes were particularly visible for the smallest values 
of wave-vectors. 

Fig.  \ref{wqfig} shows how the dispersion curve varies as the temperature increases
from $T=0.774T_c$ to $T_c$. The dispersion curves illustrated here are for the
[100] direction and are plotted up to $q=\pi/2$, which corresponds to one half of the 
Brillouin zone. As mentioned before, for larger values of $q$ the Lorentzian in 
Eq. (\ref{lorentz}) did not yield good fits to the lineshapes; however, the spin-wave
positions could still be directly read off the graphs, but with larger error bars.
For our present purpose of observing
the approach to the critical region as the temperature is raised from below $T_c$, it 
suffices to consider the dispersion curve for wave-vectors up to $q=\pi/2$.  
Well below $T_c$, the dispersion relation is linear for small $q$; as $T\to T_c$, it
changes gradually from linear to a power-law behavior of the form 
\beq
\omega_s=A_s q^x.
\label{wsfita0}
\eeq
For $T=0.9T_c$ and $L=60$ a fit including the smallest five $q$ values of the dispersion curve
to Eq. (\ref{wsfita0}) yielded $x= 1.017\pm 0.003$. As we probed further away from the Brillouin 
zone center by including larger values of $q$ in the fit, the exponent decreased slightly. 
In order to check how sensitive the fitted exponent is to the particular form of the fitted 
function, we have performed new fits to a function which includes a quadratic term, i.e., 
\beq
\omega_s=A_s q^x + B_s q^2.
\label{wsfita0q2}
\eeq
Fitting the smallest five $q$ values of the dispersion curve to a function of the form given 
by Eq. (\ref{wsfita0q2}) yielded an exponent $x=1.020\pm 0.003$, which is in good agreement
with the value obtained from the previous fit. When larger values of $q$ in the dispersion curve
were included in the fits, Eq. (\ref{wsfita0q2}) tended to yield smaller $\chi^2$'s per degree of 
freedom than Eq. (\ref{wsfita0}).
The dispersion curve for $T=T_c$ and $L=60$ fitted to Eq. (\ref{wsfita0}) yielded an exponent 
of $x=1.38\pm 0.01$ when the smallest 12 values of $q$ were included in the fit. As the larger
$q$ were excluded from the fit, the exponent increased slightly, tending towards $x\simeq 1.40$. 
When only
the smallest few values of $q$ were included in the fit, the exponent decreased again, reflecting
the fact that as we probed correlations between spins separated by larger distances (or 
equivalently, smaller $q$) the finite size of the lattice (and thus of the correlation length)
is revealed, showing that the system is not at criticality. Hence, the exponent $x$ decreases
towards unity. On the other hand, large values of $q$ correspond to short distance (in the direct 
lattice space) spin-spin correlations, and the correlation length is much larger than
the distance probed. One would thus expect the critical behavior of the system to be manifest,
and this is indeed consistent with what we obtained.
Our results at $T_c$ are in agreement with the recent experiment  \cite{coldea} which obtained
an exponent $x=1.43\pm 0.04$ when the dispersion curve at $T_c$ was fitted to a power-law
relation of the form given in Eq. (\ref{wsfita0}). 
The solid lines in Fig. \ref{wqfig} are fits to Eq. (\ref{wsfita0q2}); in general, these fits
gave lower values of $\chi^2$ per degree of freedom than fits to Eq. (\ref{wsfita0}).

In the critical region, dynamic scaling theory predicts that the half-width of spin-wave
peaks behaves as a power-law, $\Gamma_2 \sim q^{1.5}$ \cite{dynscalhh}, whereas for the
hydrodynamic regime the prediction from hydrodynamic theory is $\Gamma_2 \sim q^{2}$ 
\cite{hydrohh}.  
The half-width of the spin-wave peaks at $T=0.9T_c$ and $L=60$ from our simulations is shown in 
Fig. \ref{hwq0p9}. We observed a crossover  from  $\Gamma_2 =(0.401\pm 0.004) q^{1.46\pm 0.06}$ 
for larger values of wave-vector to $\Gamma_2 =(0.48\pm 0.02) q^{1.86\pm 0.05}$ 
for small values of $q$.
The behavior for relatively large wave-vectors is in agreement with dynamic scaling theory
and with the recent experiment \cite{coldea}. The exponent we obtained by fitting only small 
values of $q$ is close to the hydrodynamic prediction. Thus, for the spin-wave half-width 
we have observed a crossover of exponents associated with two different regimes, namely, the
critical and the hydrodynamic regions. This crossover is similar to the one observed in the 
dispersion curve at $T_c$, discussed above. For $T=T_c$ and
$L=60$ the spin-wave half-width also had a
power-law behavior which varied from approximately $q^{1.2}$ when the
12 smallest values of $q$ were included to approximately $q^{1.4}$ when only the smallest
five wave-vectors were included in the fit. In their recent experiment, Coldea {\it et al}
\cite{coldea} obtained $\Gamma_2 = D q^{1.41\pm 0.05}$ for the temperature range 
$0.77T_c\leq T < T_c$, and the coefficient $D$ increased with increasing temperature. 

As in the experiments, the dynamic structure factors from our simulations had central peaks 
(zero frequency transfer peaks) for $T\leq T_c$.
In contrast, renormalization-group theory predicts a central
peak in the longitudinal component of the dynamic structure factor only below $T_c$  
\cite{mazenkoTb}, and none of the theories predict a central peak at $T_c$ 
\cite{mazenkoTc,cuccoli}. For $T=0.9T_c$ and $L=60$ fitting the central peak half-width 
to the form $\Gamma_1\sim q^x$ yielded very large $\chi^2$ per degree of freedom. A 
much improved fit was obtained by using the function $\Gamma_1= A_1+B_1q^{C_1}$, which allows
for a non-zero central peak width when $q$ vanishes. In these
fits the data for the smallest value of $q$ observable in $L=60$, i.e. $n=1$, were not 
included, because of the large finite-size effects in them. The fit including data for 
$q$ corresponding to $n=2$ until $n=7$ yielded 
$A_1\simeq 0.013\pm 0.001$, $B_1\simeq 0.120\pm 0.005$ and $C_1\simeq 2.4\pm 0.2$. As we
systematically included larger values of $q$ in the fits, these parameters decreased slightly.
At $T_c$ we have also fitted the central peaks to Lorentzians, according to Eq. (\ref{lorentz});
however, these tended to yield curves with smaller amplitudes than the data,
as can be seen in Fig. \ref{lshapeTc}. Since there is no theoretical prediction 
for the central peak, we have also tried to fit them with a Gaussian form. These latter fits 
were, nevertheless, much worse than the fits with Lorentzians.

The lattice sizes that we used, namely $L=12$, $24$, $36$, $48$, and $60$, are all multiples
of $12$; thus, there are certain wave-vectors which are common to all lattice sizes. This 
is an advantage in the study of effects due to finite lattices, because it allows us
to compare lineshapes and spin-wave dispersion relations for different lattice sizes at a 
fixed value of wave-vector. At $T=0.9T_c$ we did not see a significant finite-size effect
for $L\geq 24$; however, when we superimposed lineshapes at $T_c$ for a fixed value of $q$, and 
different values of $L$, finite-size effects were noticeable for $L=24$. For the larger values
of $L$ that we used, the lineshapes were the same within the error bars.  

The dynamic critical exponent $z$ was extracted from the finite-size scaling of $\bar\omega_m$,
as described in a previous section. We started the analysis using no resolution function, or
equivalently $\delta_{\omega}=0$, and $n=1,2$. As in previous work \cite{alex}, we estimated
the lattice $L=30$ to be approximately the onset of the asymptotic-size regime. From the slope
of a  $\log(\bar\omega_m)$ vs $\log(L)$ graph fitted with four data points, namely lattice sizes
$L=30, 36, 48$, and $60$, we obtained $z=1.45\pm 0.01$ for $n=1$ and $z=1.42\pm 0.01$ for
$n=2$. In order to check the effects of the very small oscillations in $S({\bf q},\omega)$ due
to the finite cutoff time, we proceeded to estimate the value of $z$ using a resolution function
with $\delta_{\omega}$ given by Eq. (\ref{deltaom}). For the iterations, we used several initial 
values of $z^{(0)}$,
ranging from $z^{(0)}=1.31$ to $1.59$, and in all cases the exponent $z$ converged rapidly to
the final value with at most three iterations being necessary. For all the initial
values of $z^{(0)}$ that we used, we obtained an exponent $z=1.43\pm 0.01$ for $n=1$ and
$z=1.42\pm 0.01$ for $n=2$. In general, the $\chi^2$ per degree of freedom in the fits for
$n=2$ were slightly lower than for $n=1$, although it was reasonable in all cases. Our
final estimate for the dynamic critical exponent is  $z=1.43\pm 0.03$, where the 
error bar reflects the fluctuations in the different estimates of $z$. A comparison of
the characteristic frequency  $\bar\omega_m$ as a function of the lattice size $L$ for the 
analysis with and without a resolution function is shown in Fig. \ref{wmL}. For the former
case, the graph shown corresponds to the converged values of $z$ for both $n=1$ and $2$.  

\subsection{Comparison with experiment}

In this section we compare our results with the recent neutron scattering experiment by Coldea 
{\it et al} \cite{coldea}. Before proceeding with the direct comparison, it is necessary to 
clarify the units and possible normalization factors between simulation and experiment.

The neutron scattering experiment was done on RbMnF$_3$,  which can be described by a quantum 
Heisenberg Hamiltonian of the form
\beq
{\cal H} = J^{exp} \sum_{<{\bf rr'}>} {\bf S_r}^Q\cdot {\bf S_{r'}}^Q,
\label{hamq}
\eeq
where ${\bf S_r}^Q$ are spin operators with magnitude $|{\bf S_r}^Q|^2=S(S+1)$ and the interaction
strength between pairs of nearest-neighbors was determined experimentally \cite{windsor} to be
$J^{exp}=(0.58\pm 0.06)$ meV . In contrast, our 
simulations were performed on a classical Heisenberg Hamiltonian given in Eq. (\ref{ham}).
However, quantum Heisenberg systems 
with large spin values ($S \geq 2$) have been shown to behave as classical Heisenberg 
systems, where the spins are taken to be vectors of magnitude $\sqrt{S(S+1)}$ with the same
interaction strength between pairs of nearest neighbors as in the quantum system \cite{classqu}. 
In our simulations the spins were taken to be vectors of unit length. Hence, to preserve the
Hamiltonian the interaction strength $J$ from our simulation has to be normalized according to
\beq
J=J^{exp}S(S+1).
\label{jjexp}
\eeq 
Although this choice of normalization for spin vectors and the interaction strength leaves the 
Hamiltonian unchanged, it does modify the equations of motions.
The dynamics of the classical system so defined is the same as the
quantum system defined by the Hamiltonian in Eq. (\ref{hamq}) if one rescales the time, or
equivalently, the frequency transfer. We obtain
\beq
\omega^{exp}=J^{exp}\sqrt{S(S+1)}\; \frac{w}{J},
\label{wwexp}
\eeq
where $\omega^{exp}$ is the frequency transfer in the quantum system, measured experimentally,
and $w/J$ is the frequency transfer in units of J from our simulations. 
Parenthetically, we note that the critical temperature of the classical
Hamiltonian in Eq. (\ref{ham}) has been determined from simulations to be  
$T_c=1.442929(77)J$ \cite{kunTc}. Using the normalization for the interaction strength $J$ 
given in Eq. (\ref{jjexp}) and the experimental value $J^{exp}=(6.8\pm 0.6)$K \cite{windsor}
we get $T_c=(85.9\pm 7.6)$K, where the 9 percent error comes from
the uncertainty in $J^{exp}$. The experimental value of the critical temperature is around $83$K
which is well within the error bars. 

Due to detailed balance, neutron scattering experiments measure the dynamic structure factor
multiplied by a temperature and frequency dependent population factor 
\cite{coldea,collins,lovesey}. This factor does not appear in the simulations of the 
classical system for which the dynamic structure factor is computed 
directly. For the comparison, we removed the population factor from the experimental data.

The finite divergence of the neutron beam gives rise to a resolution function which is
usually approximated by a Gaussian in the $4$-dimensional energy and wave-vector space. In the 
experiment \cite{coldea}, the measured resolution width along the energy axis was
$0.25$ meV (full-width at half-maximum) for incoherent elastic scattering. In order to directly
compare our results with the experiment, we convoluted the lineshapes from our
simulation with a Gaussian resolution function in energy with the experimental value of
full-width at half-maximum, normalized according to Eq. (\ref{wwexp}). The standard deviation
$\delta_{\omega}$ thus obtained for the Gaussian resolution function 
was $0.0619$ in units of $J$. As a test of the effects of the resolution function in the
wave-vector space, we have also convoluted our lineshapes with a 3-dimensional Gaussian function
where, for simplicity, we have taken the resolution width in the three wave-vector components 
to be the same, and equal to the average of the experimental resolution in the longitudinal and 
transverse directions. The effect of this convolution was found to be negligible; thus the
lineshapes used in the comparisons shown below do not include the resolution in wave-vector.

The experiment \cite{coldea} performed constant wave-vector scans with both positive and 
negative energy transfer. The wave-vector transfer ${\bf Q}$ was measured along the [1,1,1] 
direction, around the antiferromagnetic zone center which in our notation is the $(\pi,\pi,\pi)$ 
point. Note that Ref. \cite{coldea} defines the wave-vector transfer ${\bf Q}$ in units such
that the antiferromagnetic zone center is $(0.5,0.5,0.5)$; hence, to express ${\bf Q}$ of 
Ref. \cite{coldea} in units of \AA$^{-1}$ one has to multiply it by $2\pi/a$, where $a$ 
is the cubic lattice parameter expressed in \AA. However, in the simulation 
direct lattice positions are defined in units of the lattice constant $a$; thus we obtain
wave-vectors multiplied by the constant $a$. Let us emphasize that one has to divide the 
wave-vector ${\bf Q}$ [and also $q$, see Eq. (\ref{qpbc})] defined in this paper by $2\pi$ in
order to express it in the units used in Ref. \cite{coldea}. In the experiment, 
measurements were taken for wave-vectors ${\bf Q}=(\pi+q,\pi+q,\pi+q)$, with $q=2\pi(0.02),$ 
$2\pi(0.04),$..., $2\pi(0.12)$, but unfortunately these values of $q$ are not all observed 
in our simulations with the particular lattice sizes that we used. For instance, with a 
lattice size $L=60$ we observe wave-vectors with $q=2\pi(0.01666...),$ $2\pi(0.03333...)$,... 
and so on, according to Eq. (\ref{qpbc}). Thus, in order to directly compare the lineshapes 
from the simulation with the experimental ones, it was necessary to interpolate our results 
to obtain the same $q$ values of the experiment. This was done by first fitting our lineshapes 
with a Lorentzian form, as given in Eq. (\ref{lorentz}). Since the parameters $B$, $\Gamma_2$ 
and $\omega_s$ obtained from these fits behave as power-laws of $q$, we linearly interpolated 
the logarithm of these parameters as a function of the logarithm of $q$, to obtain new 
parameters for the lineshapes corresponding to those values of $q$ observed in the experiment. 
We estimated the uncertainties from this procedure to be less than five percent for the parameter 
$B$, less than three percent for the spin-wave half-width $\Gamma_2$ and the spin-wave position
$\omega_s$ at $T_c$, and less than one percent for the spin-wave position $\omega_s$ at 
$T=0.9T_c$. Below $T_c$, the parameters $A$ and $\Gamma_1$ associated with the central 
peak were linearly interpolated, yielding new parameters with uncertainties of approximately
five percent. At $T_c$, the parameter $A$ was interpolated in the log-log plane (as for 
$B$, $\Gamma_2$ and $\omega_s$ discussed above), whereas $\Gamma_1$ was simply linearly 
interpolated. The uncertainties in $A$ and $\Gamma_1$ at $T_c$ were estimated to be less 
than ten percent. For $L=60$, there is one value of $q$, namely $q=2\pi(0.10)$,
which is accessible to both simulation and experiment. This was the only case for which we
did not have to interpolate in $q$. 

Below $T_c$, the simulations are mainly for $T=0.9T_c$ which unfortunately 
does not coincide with any temperature used in the experiment; however, it is very close
to $T=0.894T_c$ which is one of the temperatures for which experimental results are available.
To correct for the slight difference, we made a linear interpolation in temperature, using
our results for $L=24$ at $T=0.846T_c$ and at $T=0.9T_c$. We first fitted the lineshapes at 
these two temperatures to a Lorentzian of the form given by Eq. (\ref{lorentz}), then we
linearly interpolated the position and the amplitude of the spin-wave peak at these 
temperatures, to obtain the spin-wave position and amplitude corresponding to $T=0.894T_c$. 
For small values of $q$ we found that the frequency of the spin-wave peak at $T=0.894T_c$ 
was approximately $1.5$ percent larger than at $T=0.9T_c$ and this difference decreased for 
larger values of $q$. The spin-wave amplitude at $T=0.894T_c$ was found to be approximately 
$5$ percent larger than at $T=0.9T_c$ for small values of $q$. As for the spin-wave position, 
the difference in the amplitudes decreased for larger values of wave-vector. 

The intensity of the lineshapes in the neutron scattering experiment was measured in
counts per 15 seconds. For both temperatures $T=0.894T_c$ and $T=T_c$ the measurements for
the several wave-vectors were done with the same experimental set-up and conditions.
Therefore, the relative intensities of the lineshapes for the different wave-vectors is fixed,
and equal for both temperatures. The intensity of the lineshapes obtained in the simulation 
had to be normalized to the experimental value; however, because the relative intensities for 
different wave-vectors is fixed, we have only one independent normalization factor for all the 
wave-vectors at both temperatures. The normalization of the intensity was chosen so that
the spin-wave peak for $T=0.894T_c$ and $q=2\pi(0.08)$ from the experiment and the simulation 
matched. This same factor was used to normalize the intensities of the lineshapes corresponding 
to the remaining values of wave-vectors at $T=0.894T_c$, and for all cases at $T_c$. The factor 
used was $70$ counts/15 secs., which we multiplied to the simulated lineshapes for all values 
of $q$ at $T=0.894T_c$ and at $T_c$. 

The final lineshapes for $T=0.894T_c$, $L=60$, and several wave-vectors are 
shown in Fig. \ref{ceTp894} together with experimental lineshapes for each case.
Figs. \ref{cewqTp9}(a) and \ref{cewqTp9}(b) show respectively the comparisons of the 
dispersion curve and the spin-wave half-width from the simulation and the experiment at 
$T=0.894T_c$. The good agreement between our results and experiment can be
seen from either the direct comparison of the lineshapes, or the comparisons of the dispersion
curve and the spin-wave half-width. There is an agreement between the lineshape intensities from 
simulation and experiment over two orders of magnitude, from $q=2\pi(0.02)$ to $q=2\pi(0.10)$.
Fig. \ref{ceTc} shows the comparison of lineshapes from the simulation and the experiment for
$T=T_c$, $L=60$, and several values of $q$. The dispersion curve obtained from the simulations 
at $T=T_c$, shown in Fig. \ref{cewqTc}, is systematically larger than the experimental values. 
We would like to emphasize that the
error bars shown for the dispersion curve obtained from our simulations at $T_c$ reflect only 
the statistical errors of a best fit of the lineshapes with Eq. (\ref{lorentz}). For each
wave-vector, this fit was done with only one range of frequency; hence errors associated with 
the choice of frequency range and the quality of the fit were not taken into account. It is 
reasonable to expect that such sources of error would increase the error bars by a factor of 5.
From the direct comparison of the simulated and experimental lineshapes at $T_c$ it is difficult 
to determine the difference in the spin-wave frequencies, because the spin-wave peaks from the 
experiment are not very pronounced, and their positions have to be extracted from the fits of 
the lineshapes. As we mentioned before, the experimental data at $T_c$ were fitted to a 
functional form predicted by RNG theory plus a Lorentzian central peak. As an illustration,
one such fit is shown in Fig. \ref{ceTc}(c) for $q=2\pi(0.08)$, along with the RNG component of 
the fit and the prediction by mode-coupling theory. Finally, even though at $T_c$ the lineshape 
intensities from the simulations for small frequency transfer tended to be lower as compared to 
the experiment, the agreement is still reasonably good, considering the variation of the 
intensities over almost two orders of magnitude from $q=2\pi(0.02)$ to $q=2\pi(0.12)$.

\section{Conclusion}

We have studied the dynamic critical properties of the classical Heisenberg antiferromagnet
in a simple cubic lattice, using large-scale computer simulations. A new time integration
technique implemented by Krech {\it et al} allowed us to use a larger time integration step
and we were thus able to extend the maximum integration time to much larger values than in
previous work.  

Below $T_c$, the dispersion curves were approximately linear for wave-vectors 
well within the first Brillouin zone. Increasing the temperature towards the critical
temperature the dispersion curve became a power-law, reflecting the crossover from
hydrodynamic to critical behavior. The power-law behavior of the spin-wave half-width at
$T=0.9T_c$ also showed a crossover from critical behavior at large values of $q$ to 
hydrodynamic behavior at small values of $q$. The dynamic critical exponent was estimated
to be $z=(1.43\pm 0.03)$ which is in agreement with the experimental value of Coldea {\it et al},
and slightly lower than the dynamic scaling prediction. 

We made direct, quantitative comparison of both the dispersion curve and the lineshapes 
obtained from our simulations with the recent experimental results by Coldea {\it et al}. 
At $T=0.894T_c$ the agreement was very good. The major difference was at $T_c$ where spin-wave 
peaks from our simulations tended to be at slightly larger frequencies than the experimental 
results. Both at $T=0.894T_c$ and at $T_c$ the lineshape intensities varied over almost two
orders of magnitude from $q=2\pi(0.02)$ to $q=2\pi(0.10)$ and there was good agreement 
between the intensities from simulation and experiment over the whole range. Thus, the simple 
isotropic nearest-neighbor classical Heisenberg model is very good for describing the 
dynamic behavior of this real magnetic system, except for small differences in spin-wave 
frequencies at the critical temperature.

\begin{center}
{\bf ACKNOWLEDGMENTS}
\end{center}

We are indebted to Professor R. A. Cowley and Dr. R. Coldea for very helpful discussions
and for sending us ascii files of their data. We would also like to thank Dr. M. Krech
and Professor 
H. B. Sch\"{u}ttler for valuable discussions. Computer simulations were carried out on 
the Cray T90 at the San Diego Supercomputing Center, and on a Silicon Graphics Origin2000 
and IBM R6000 machines in the University of Georgia. This research was supported in part 
by NSF Grant No. DMR-9727714.

\vspace{2cm}

\begin{center}
{\bf FIGURE CAPTIONS}
\end{center}

\begin{figure}
%\vspace{-4cm}
\centering
\leavevmode
\epsfxsize=3.0in
\begin{center}
\leavevmode
\epsffile{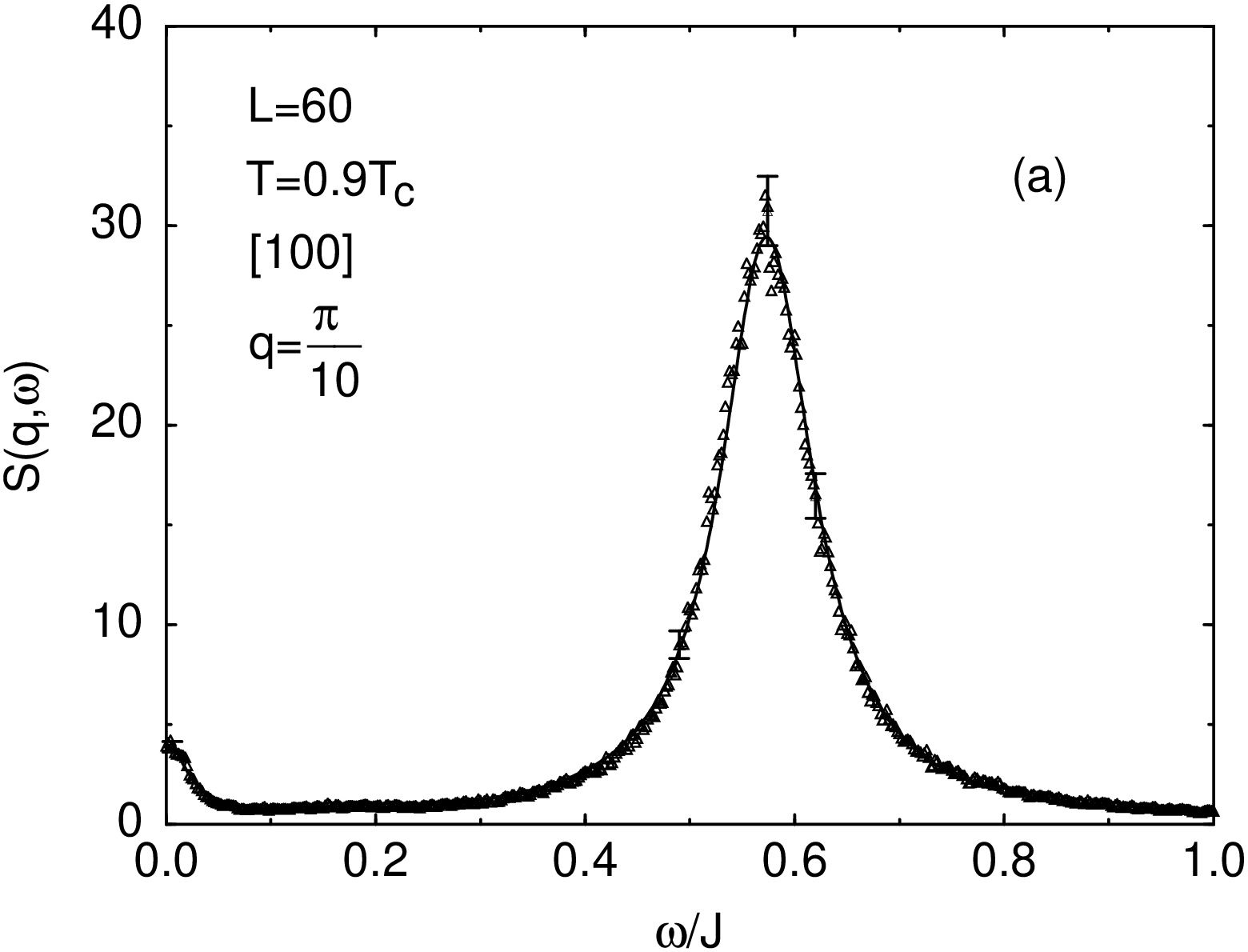}
\end{center}
%\vspace{-2cm}
\begin{center}
\leavevmode
\epsfxsize=3.0in
\epsffile{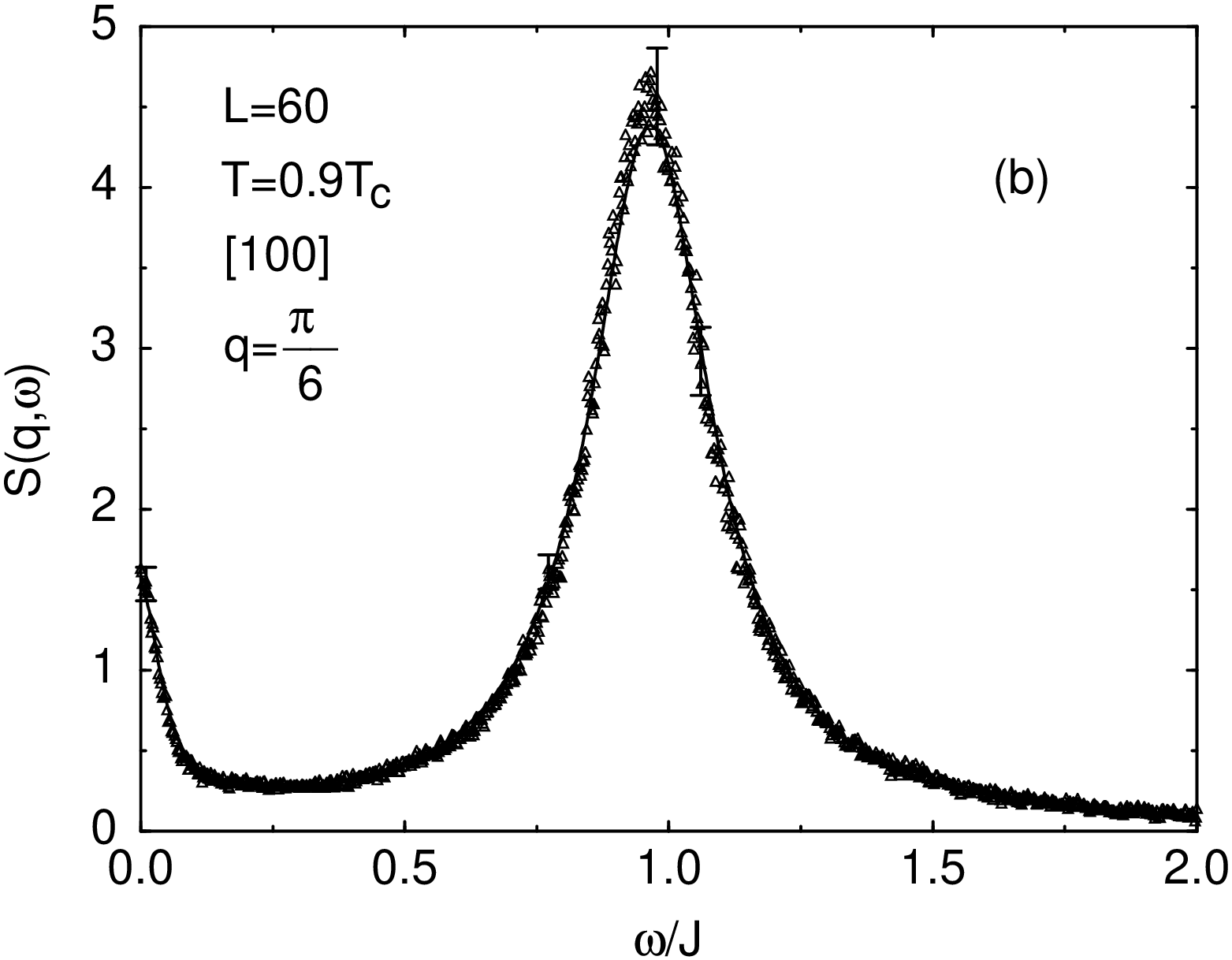}
\end{center}
\vspace{-1cm}
\begin{center}
\leavevmode
\epsfxsize=3.0in
\epsffile{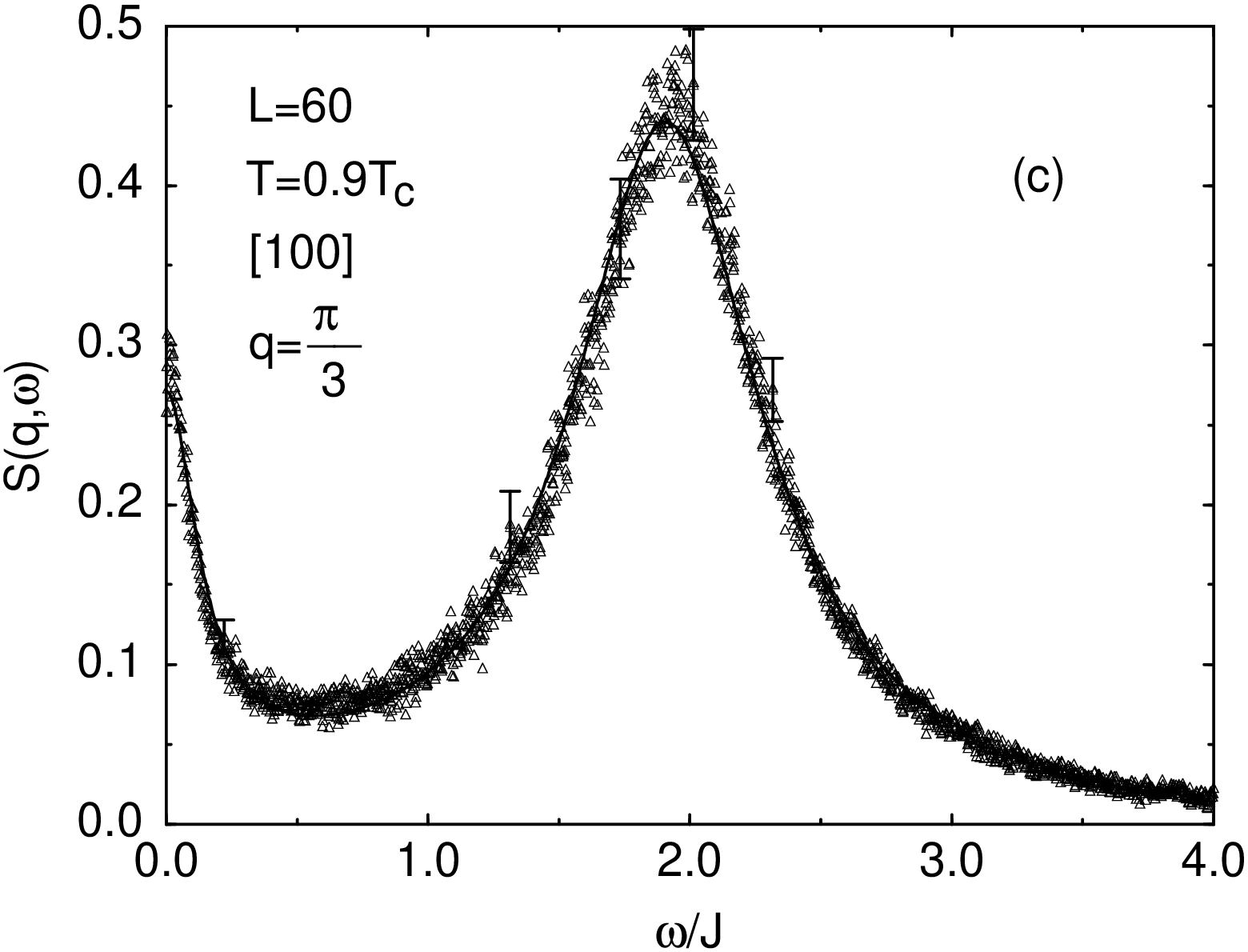}
\end{center}
%\vspace{-2cm}
\caption{}
\label{lshape09}
\end{figure}

\begin{figure}
%\vspace{-4cm}
\centering
\leavevmode
\epsfxsize=3.0in
\begin{center}
\leavevmode
\epsffile{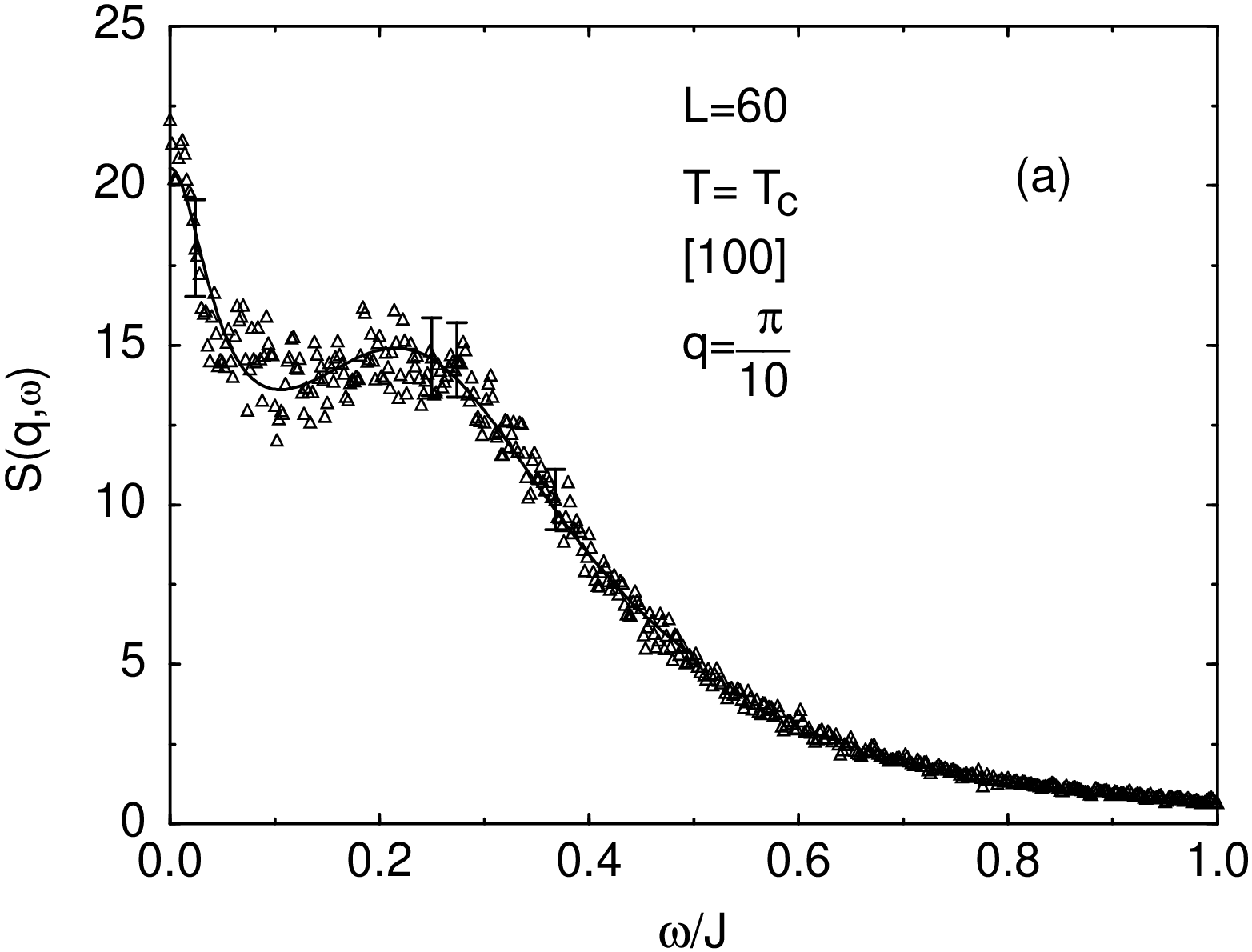}
\end{center}
\vspace{-2cm}
\begin{center}
\leavevmode
\epsfxsize=3.0in
\epsffile{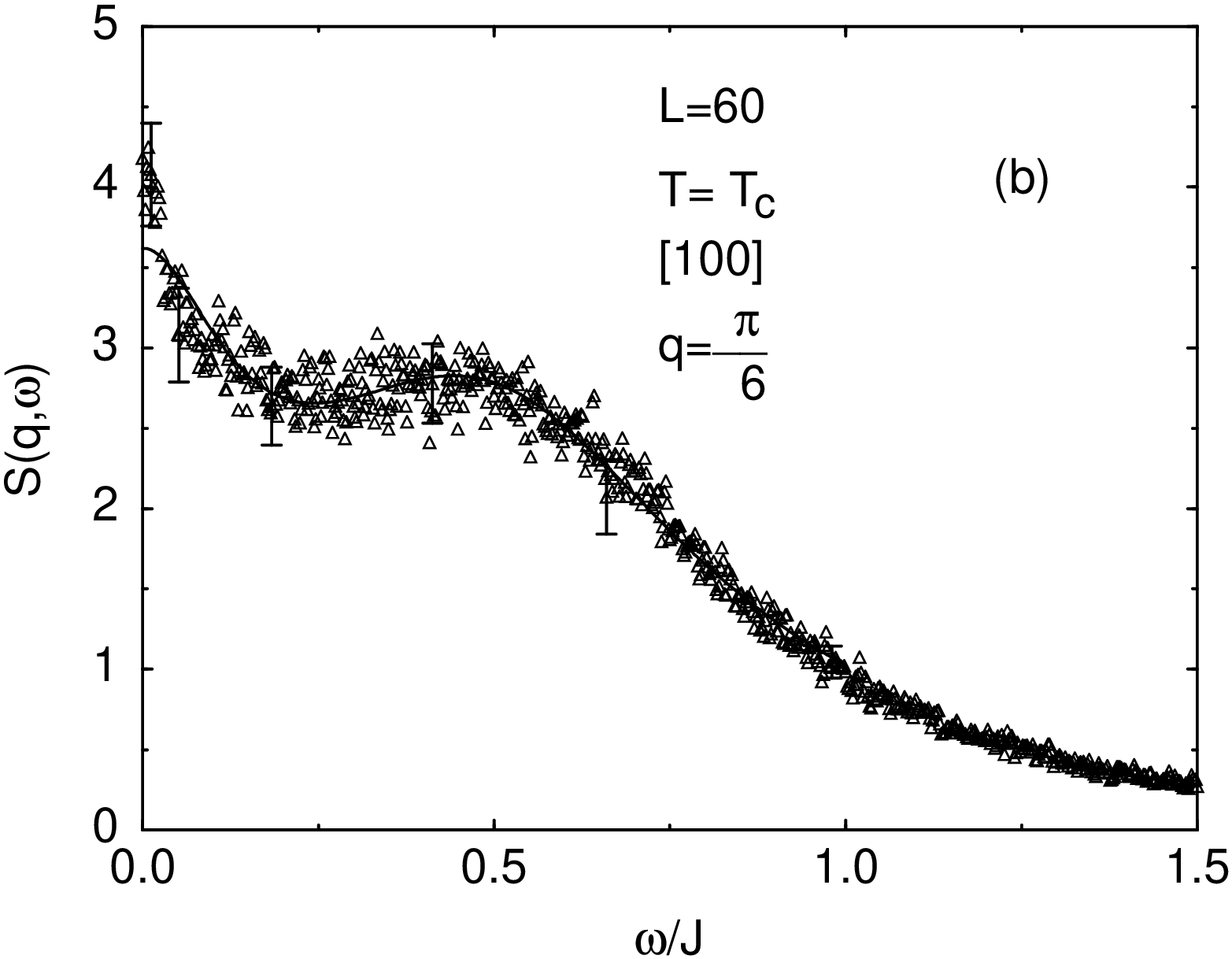}
\end{center}
%\vspace{-2cm}
\caption{}
\label{lshapeTc}
\end{figure}

\begin{figure}
\centering
\leavevmode
\epsfxsize=4.0in
\epsffile{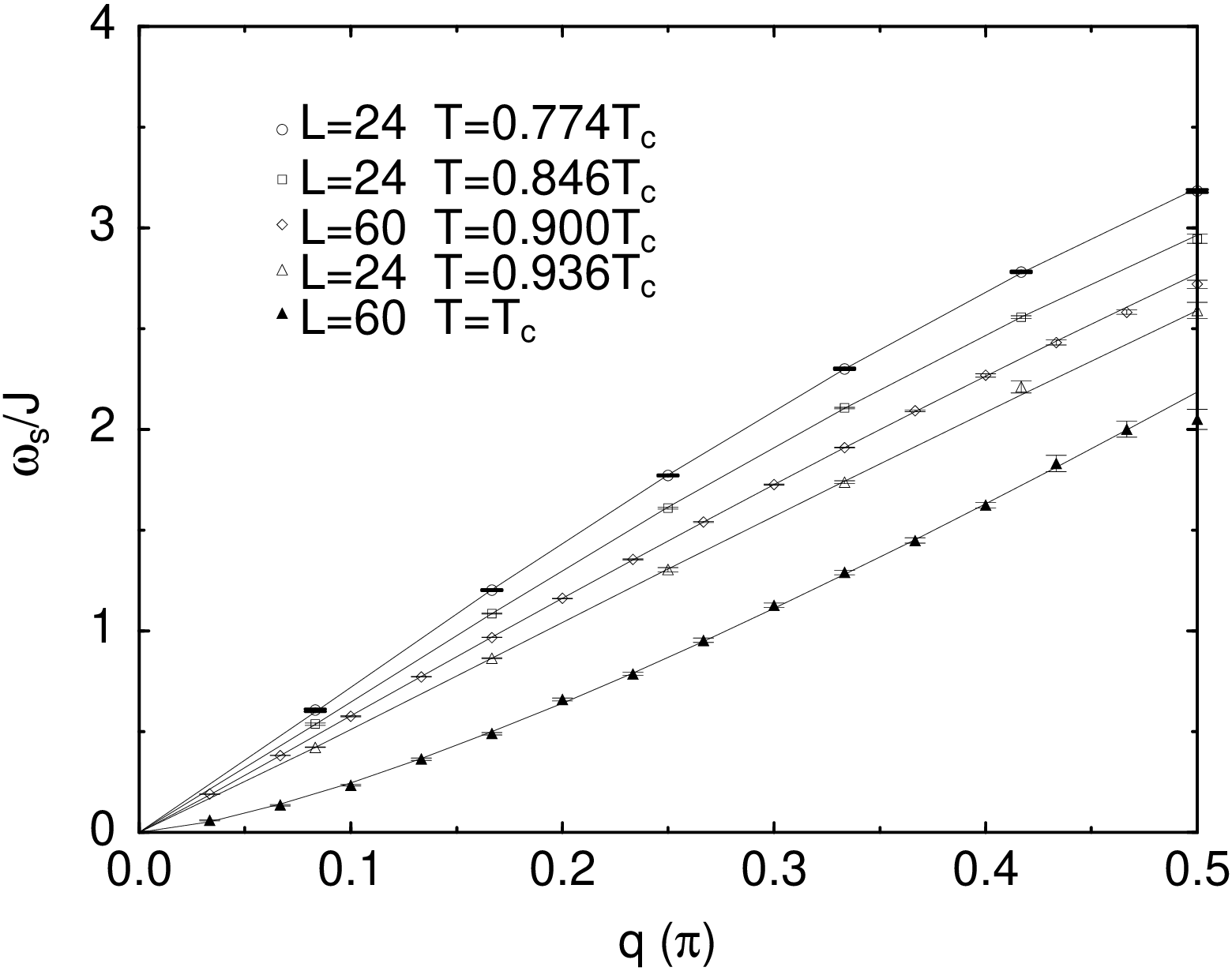}
\caption{}
\label{wqfig}
\end{figure}

\begin{figure}
\centering
\leavevmode
\epsfxsize=4.0in
\epsffile{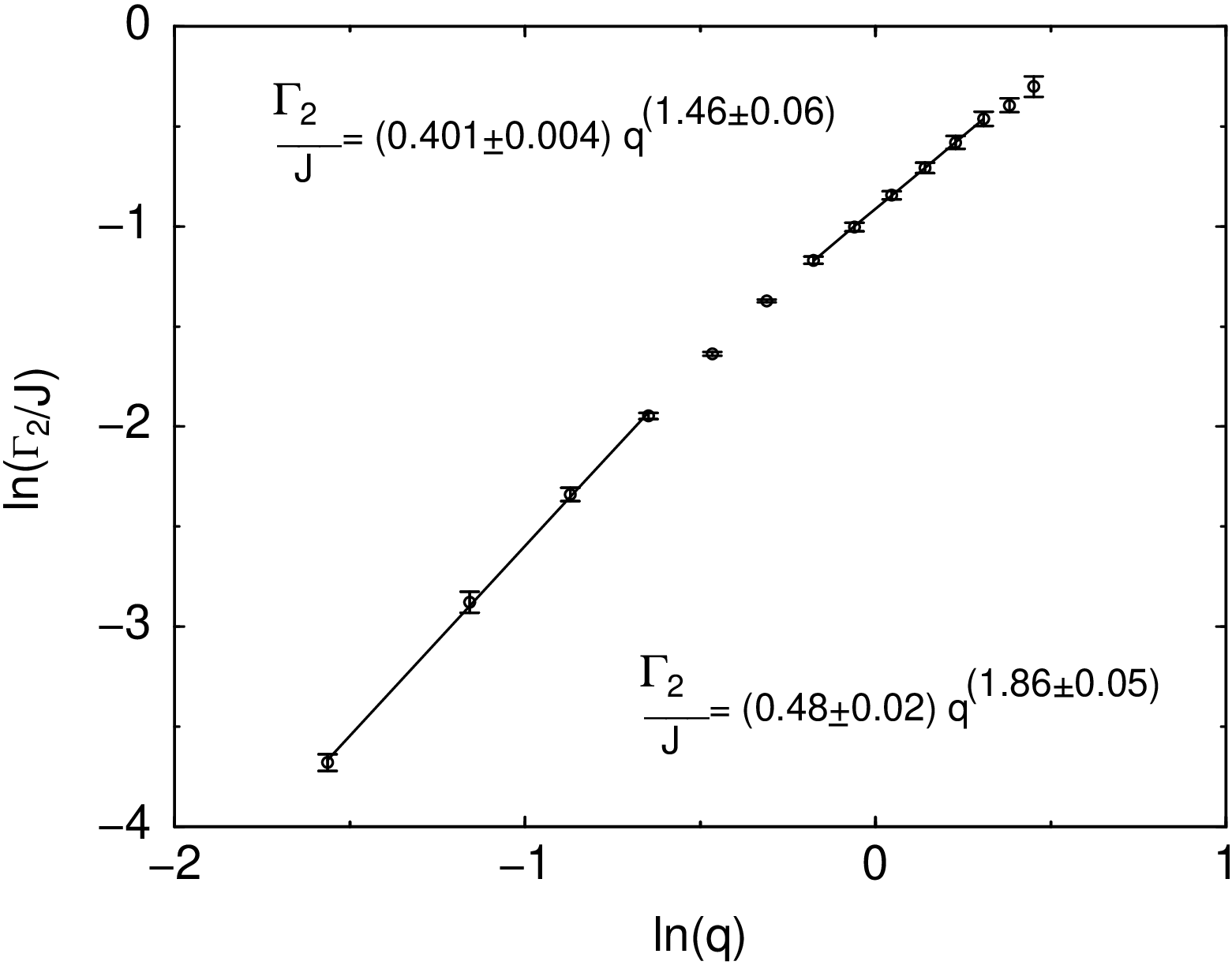}
\caption{}
\label{hwq0p9}
\end{figure}

\begin{figure}
\vspace{-1 cm}
\centering
\leavevmode
\epsfxsize=3.5in
\epsffile{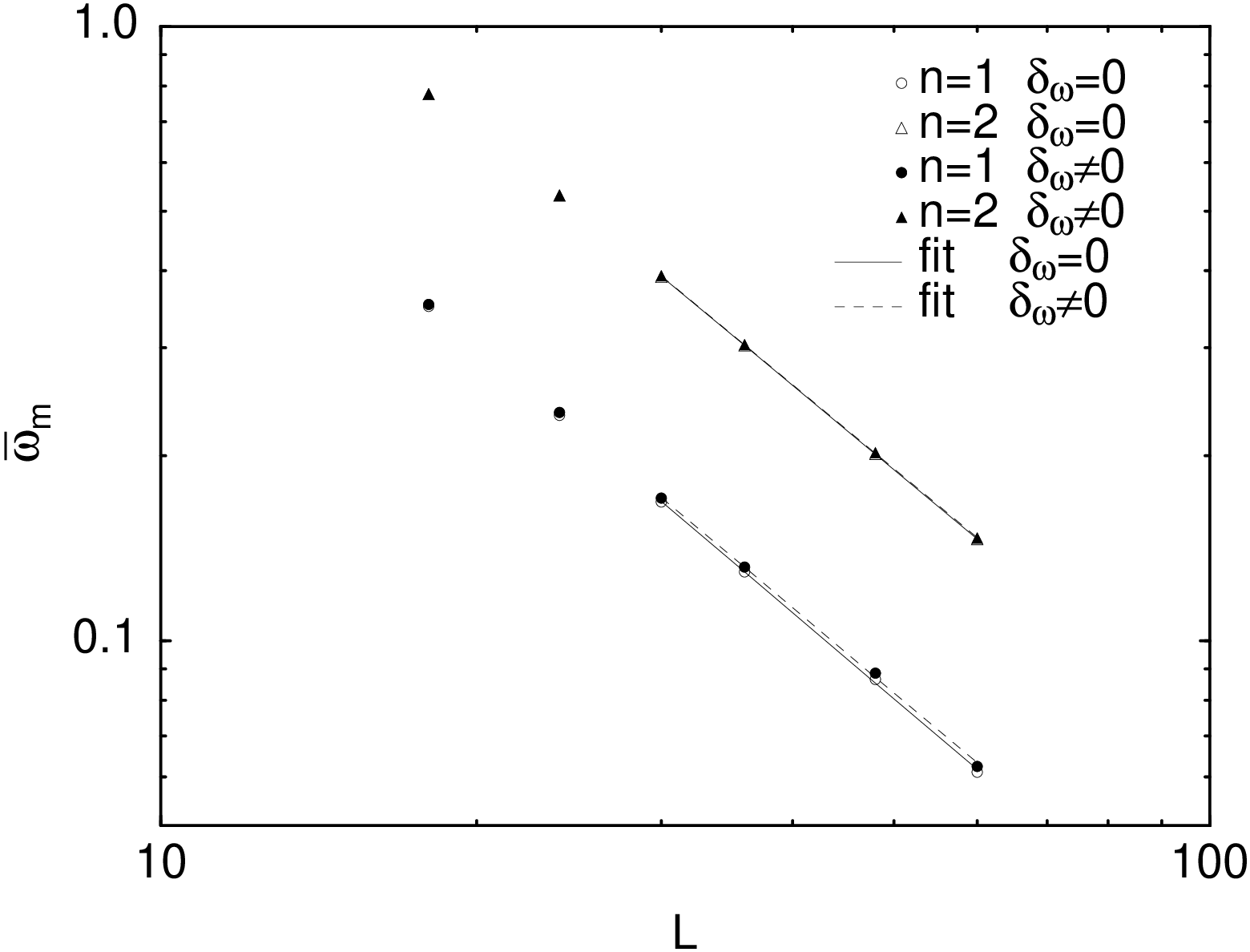}
\caption{}
\label{wmL}
\end{figure}

\begin{figure}
%\vspace{-4cm}
\centering
\leavevmode
\epsfxsize=3.0in
\begin{center}
\leavevmode
\epsffile{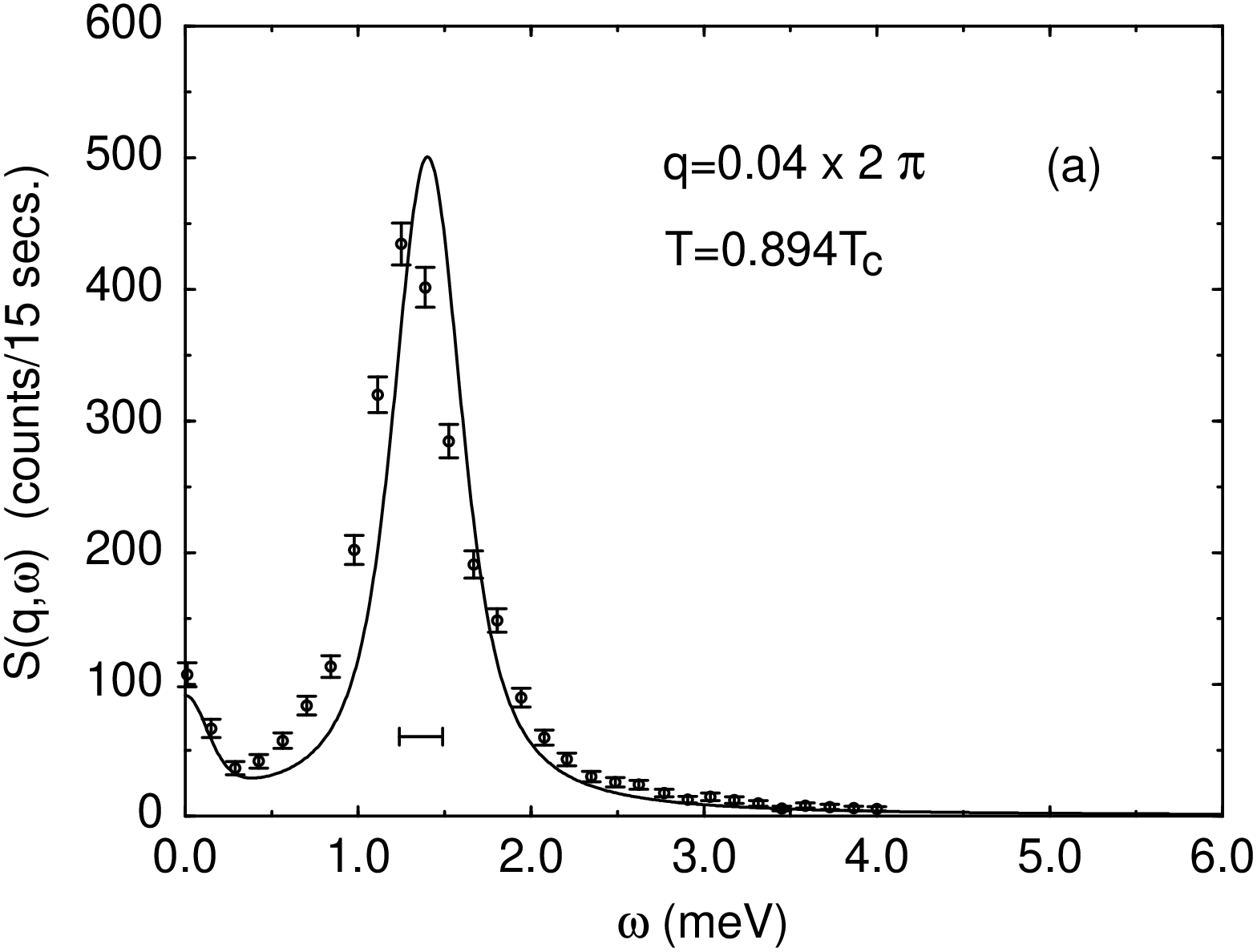}
\end{center}
\vspace{-3cm}
\begin{center}
\leavevmode
\epsfxsize=3.0in
\epsffile{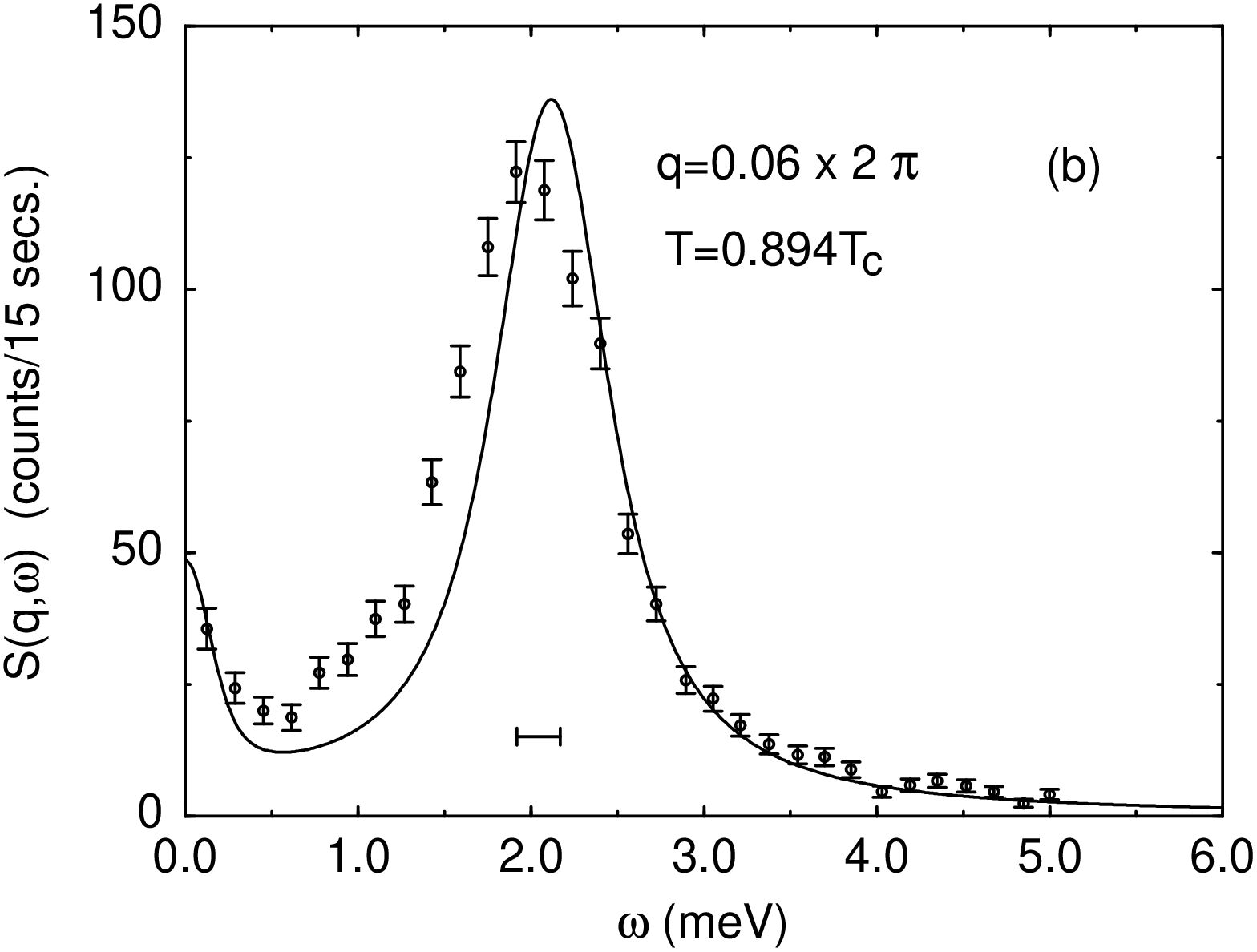}
\end{center}
\vspace{-3cm}
\begin{center}
\leavevmode
\epsfxsize=3.0in
\epsffile{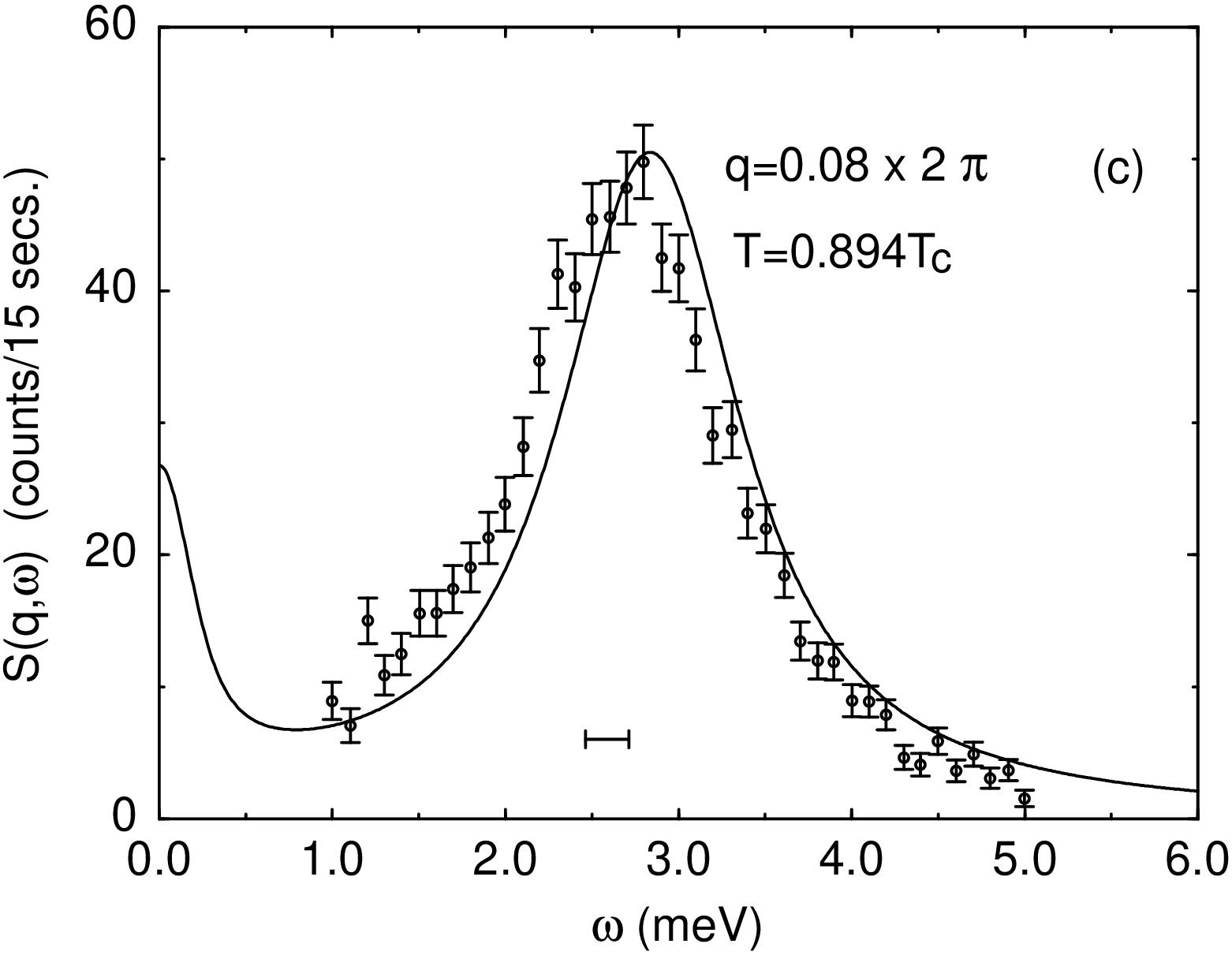}
\end{center}
\vspace{-3cm}
\begin{center}
\leavevmode
\epsfxsize=3.0in
\epsffile{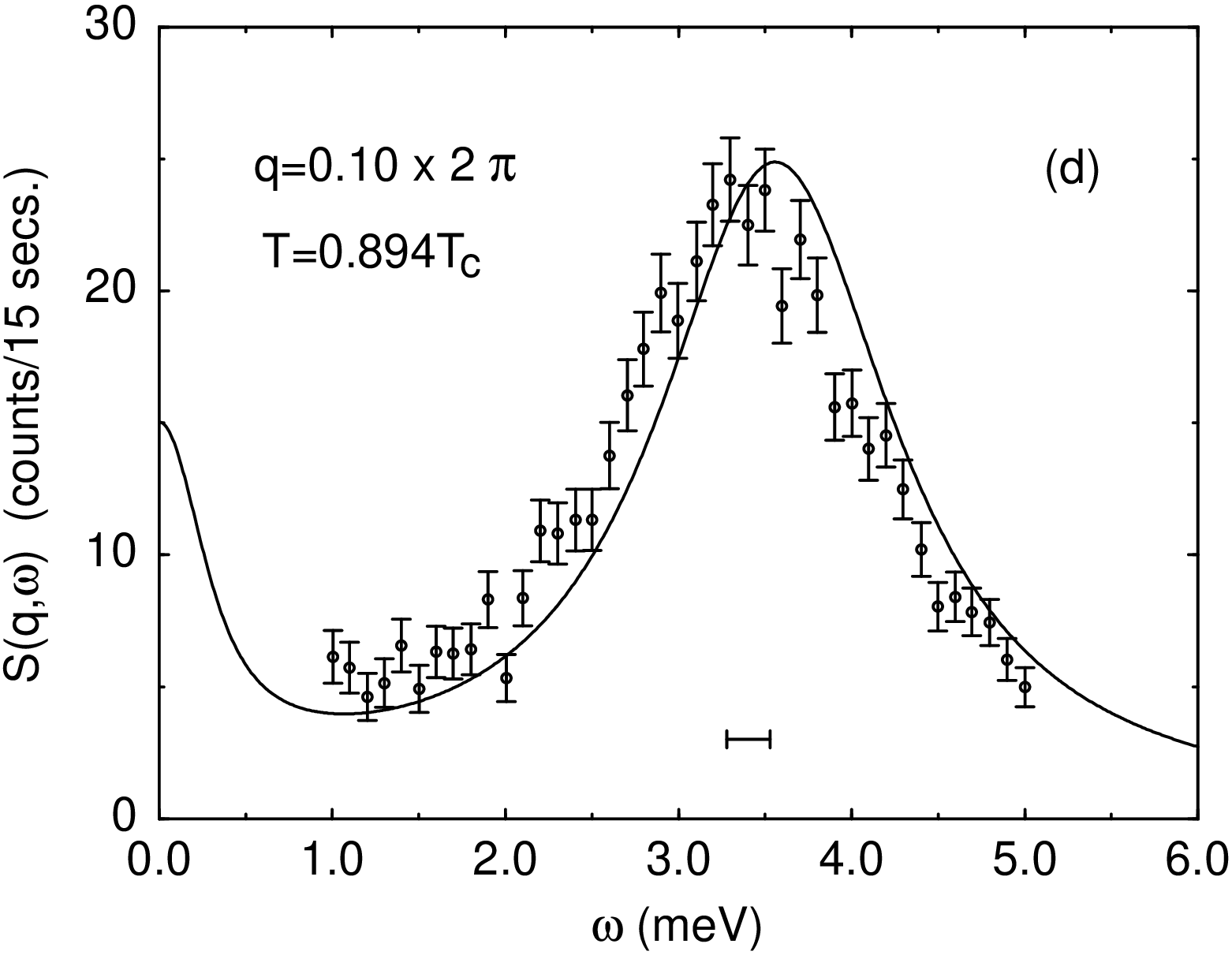}
\end{center}
%\vspace{-2cm}
\caption{}
\label{ceTp894}
\end{figure}

\begin{figure}
%\vspace{-4cm}
\centering
\leavevmode
\epsfxsize=3.2in
\begin{center}
\leavevmode
\epsffile{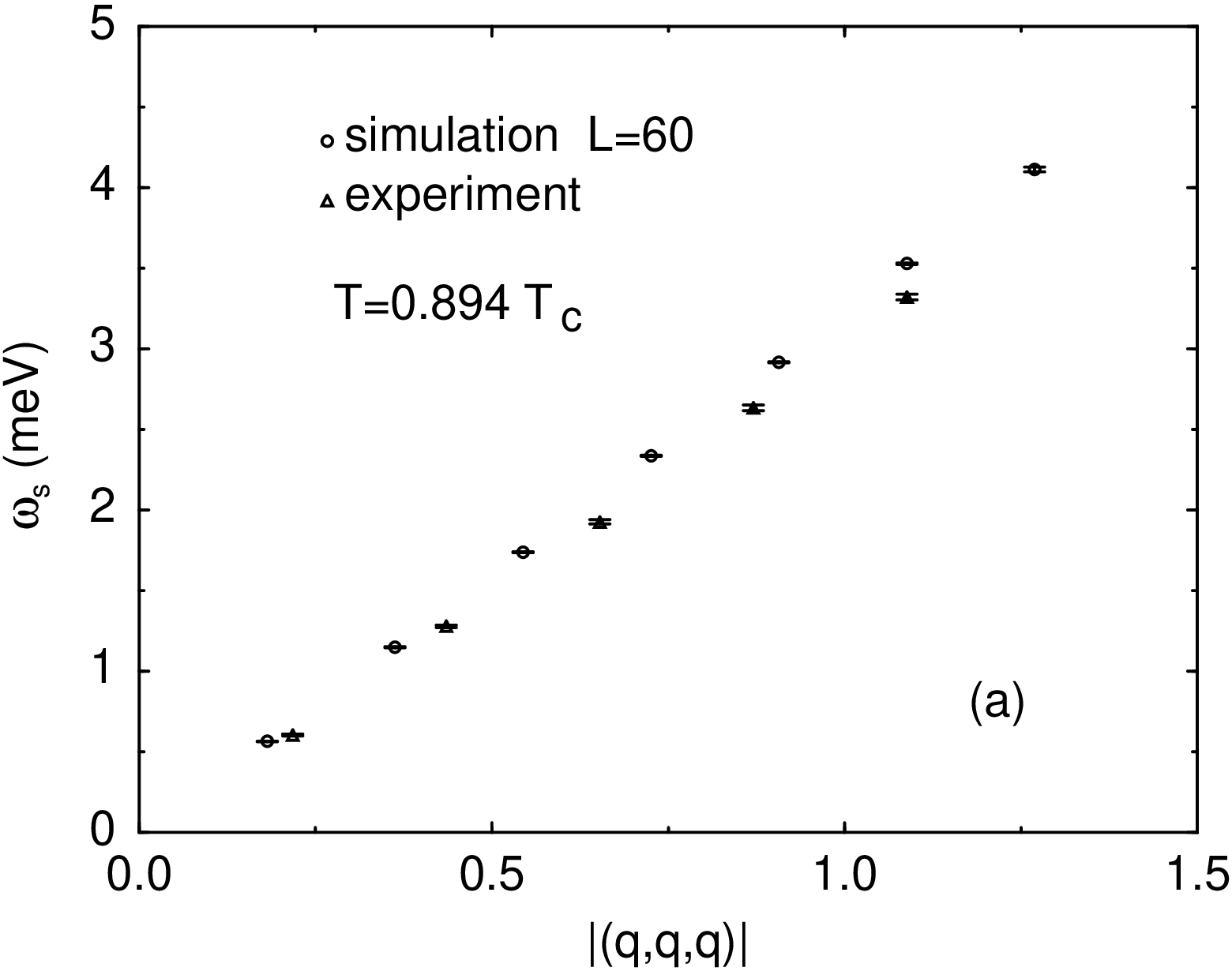}
\end{center}
\vspace{-3cm}
\begin{center}
\leavevmode
\epsfxsize=3.2in
\epsffile{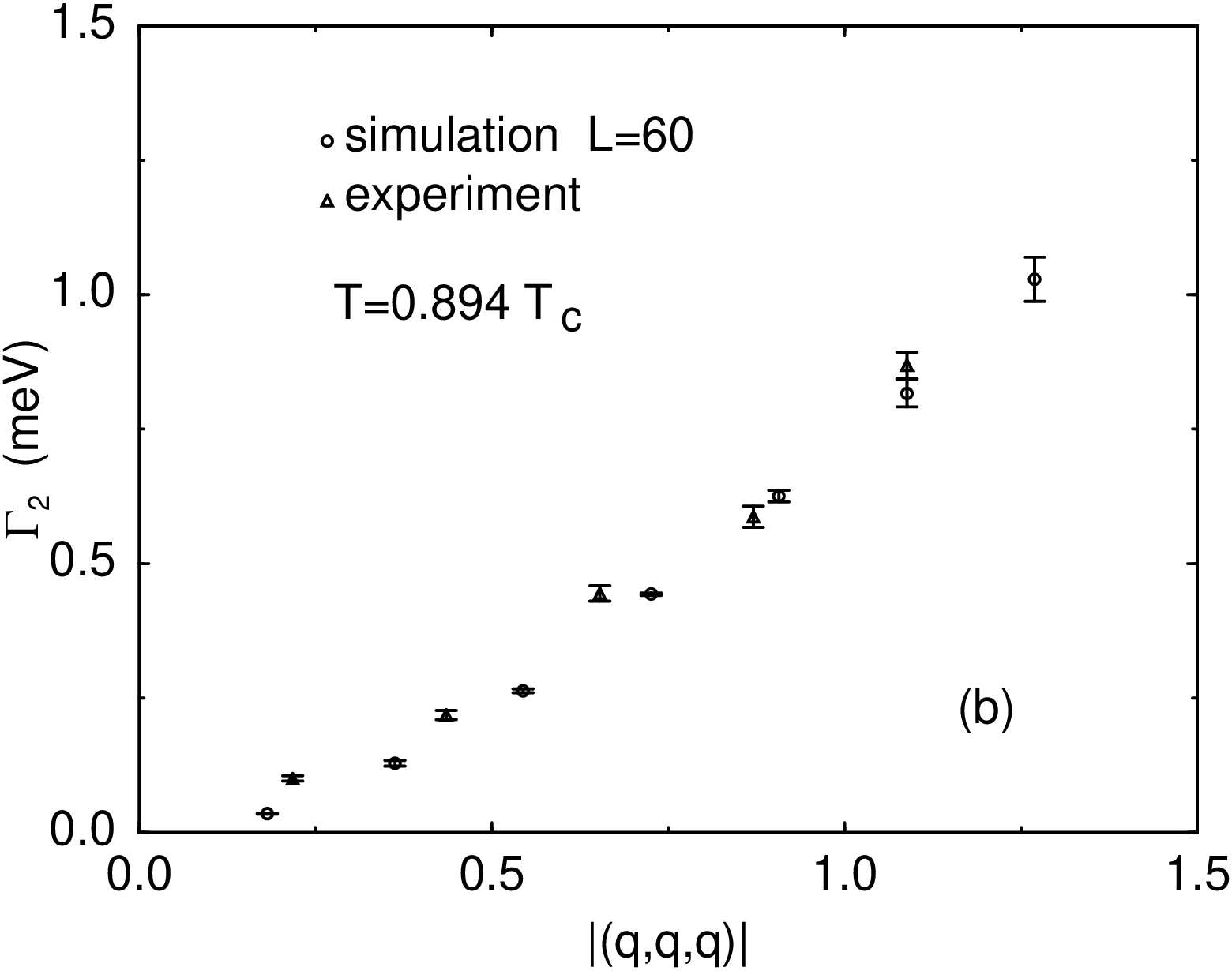}
\end{center}
%\vspace{-2cm}
\caption{}
\label{cewqTp9}
\end{figure}

\begin{figure}
%\vspace{-4cm}
\centering
\leavevmode
\epsfxsize=3.2in
\begin{center}
\leavevmode
\epsffile{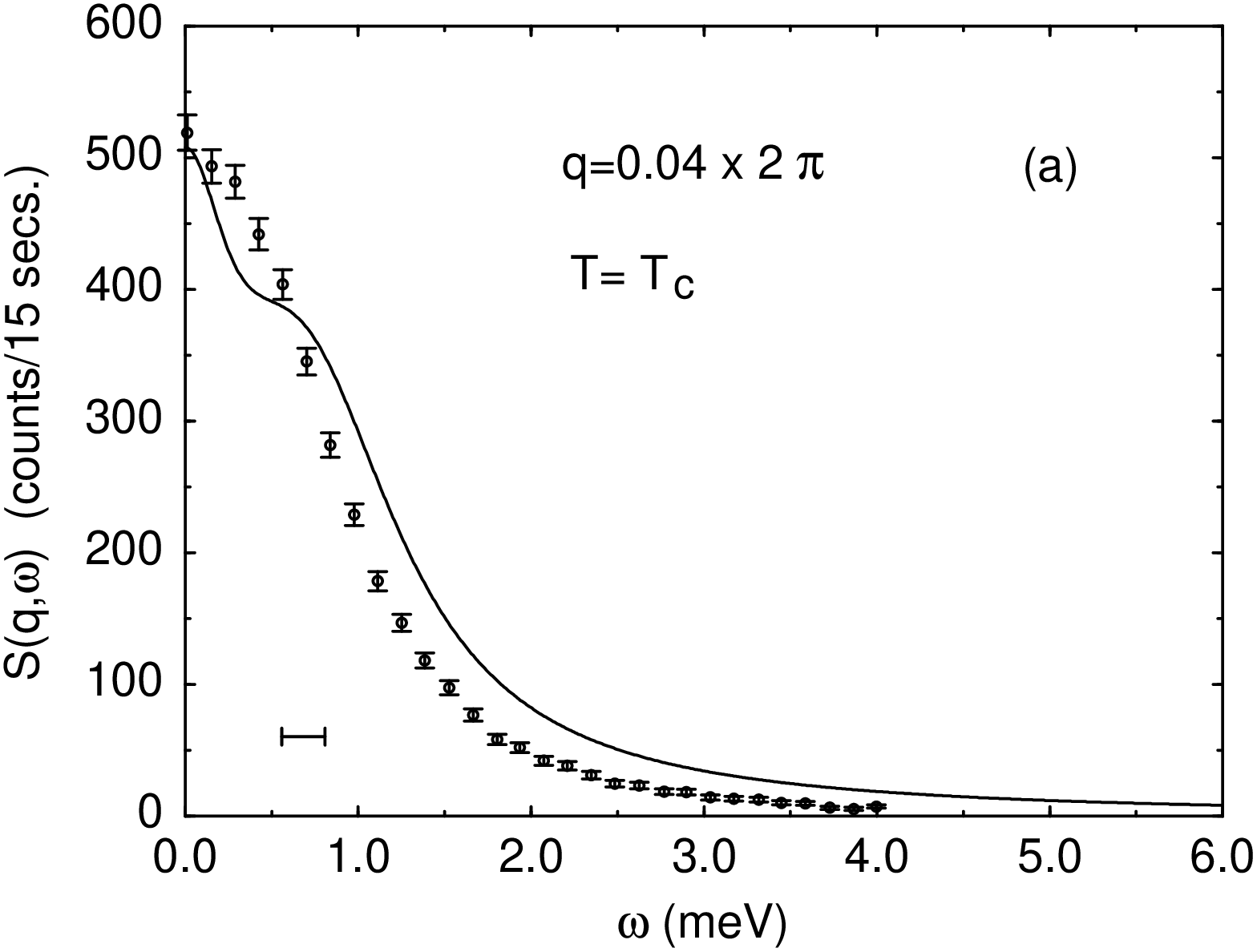}
\end{center}
\vspace{-3cm}
\begin{center}
\leavevmode
\epsfxsize=3.2in
\epsffile{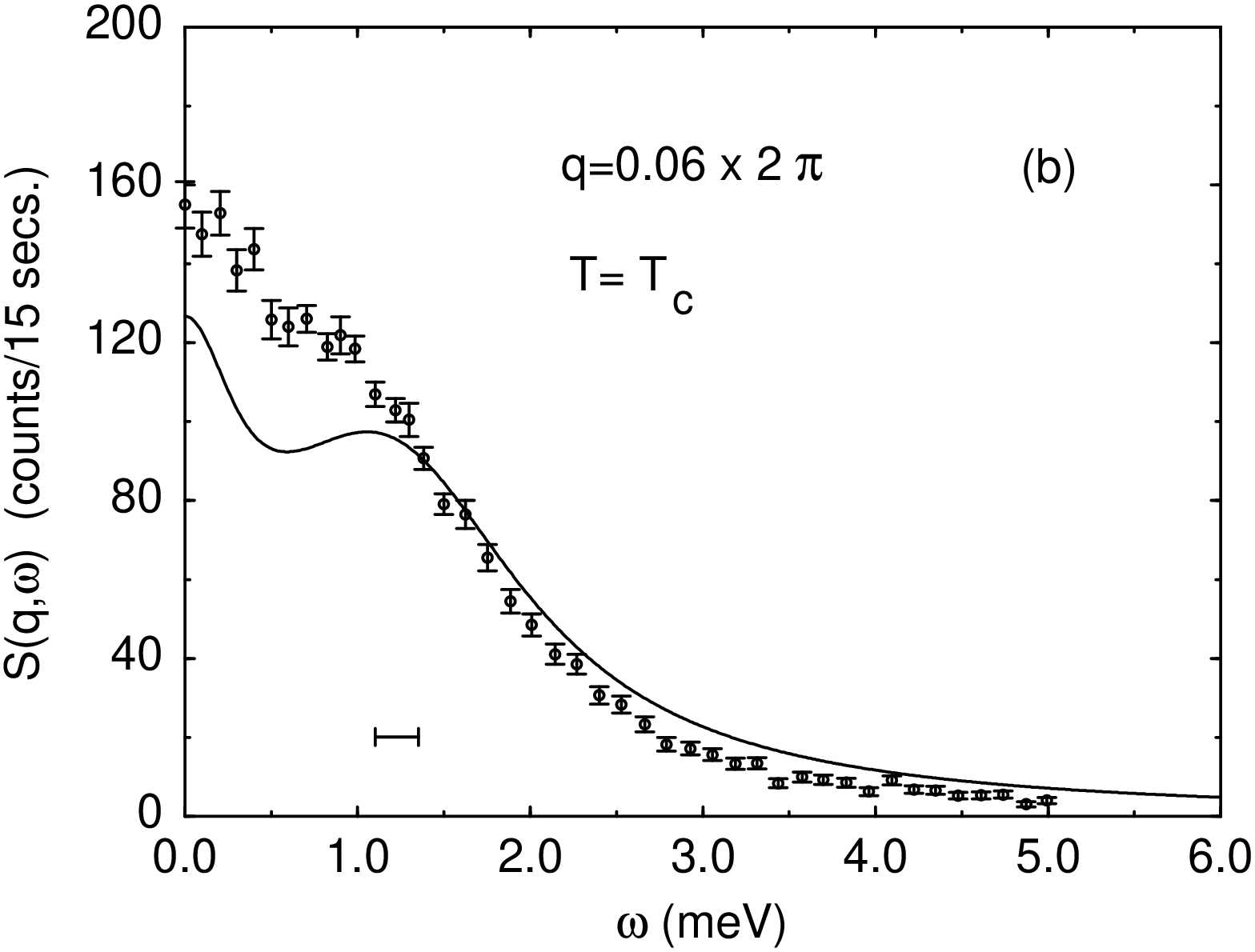}
\end{center}
\vspace{-2cm}
\begin{center}
\leavevmode
\epsfxsize=3.2in
\epsffile{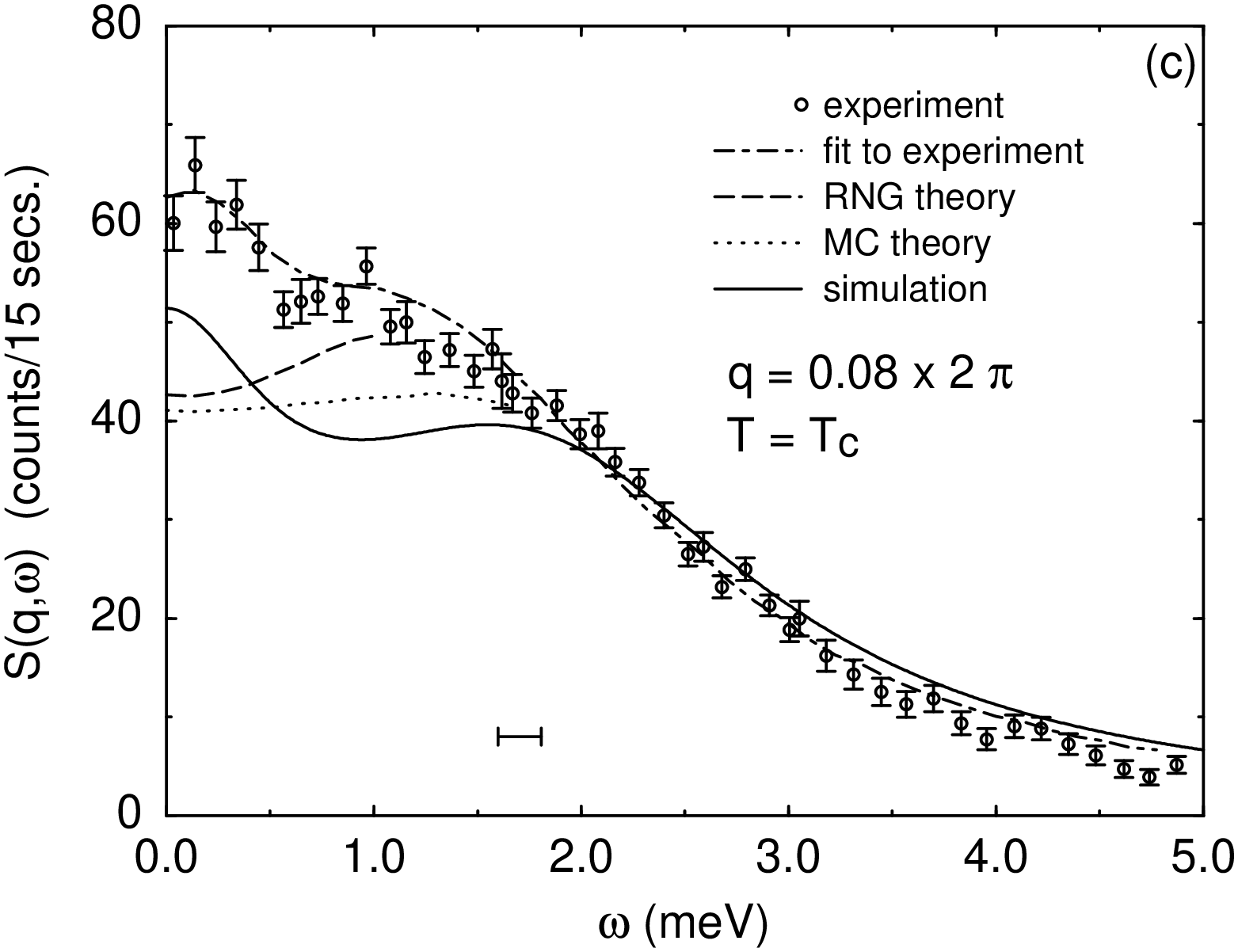}
\end{center}
\vspace{-3cm}
\begin{center}
\leavevmode
\epsfxsize=3.2in
\epsffile{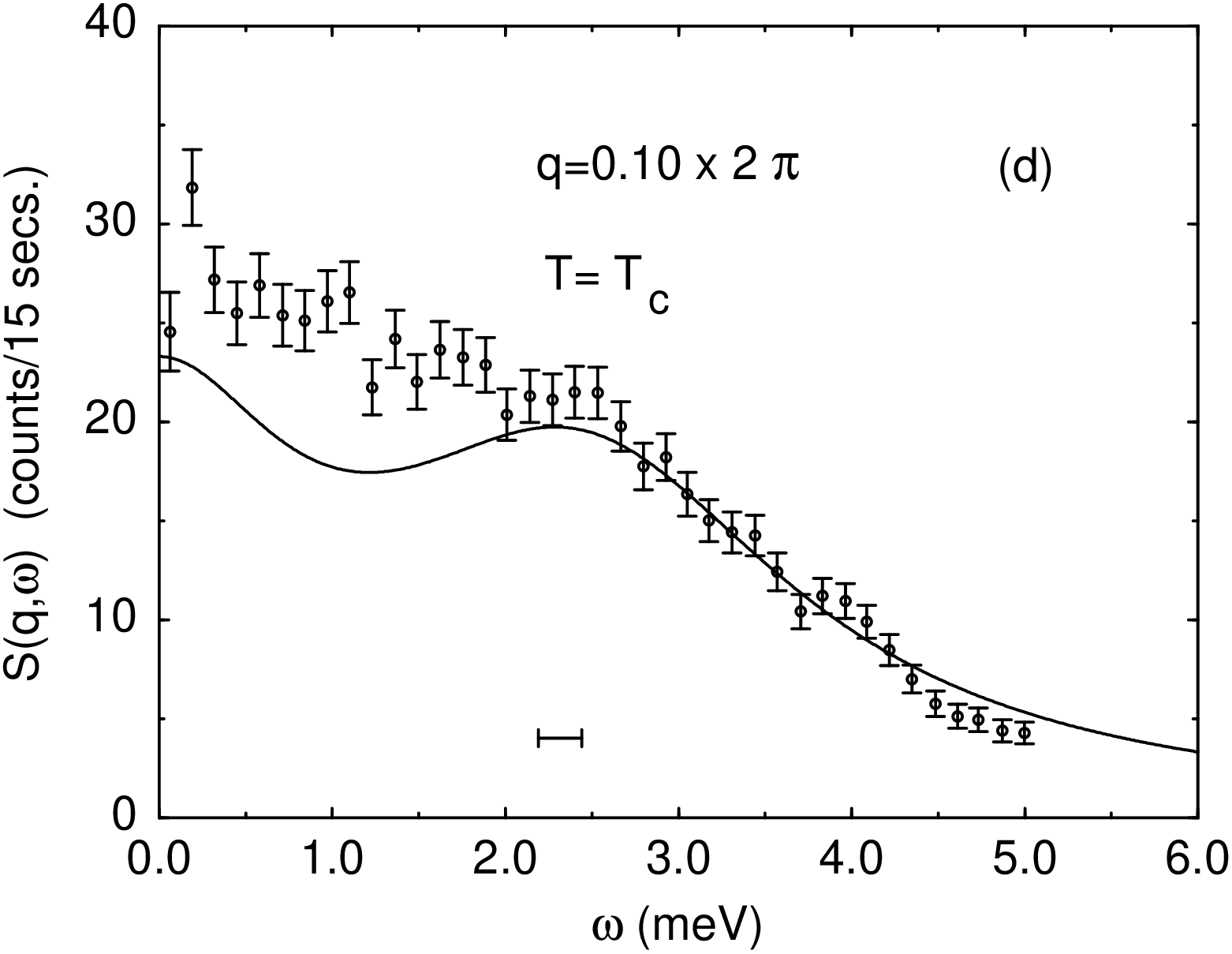}
\end{center}
\vspace{-1cm}
\caption{}
\label{ceTc}
\end{figure}

\begin{figure}
%\vspace{-4cm}
\centering
\leavevmode
\epsfxsize=3.2in
\begin{center}
\leavevmode
\epsffile{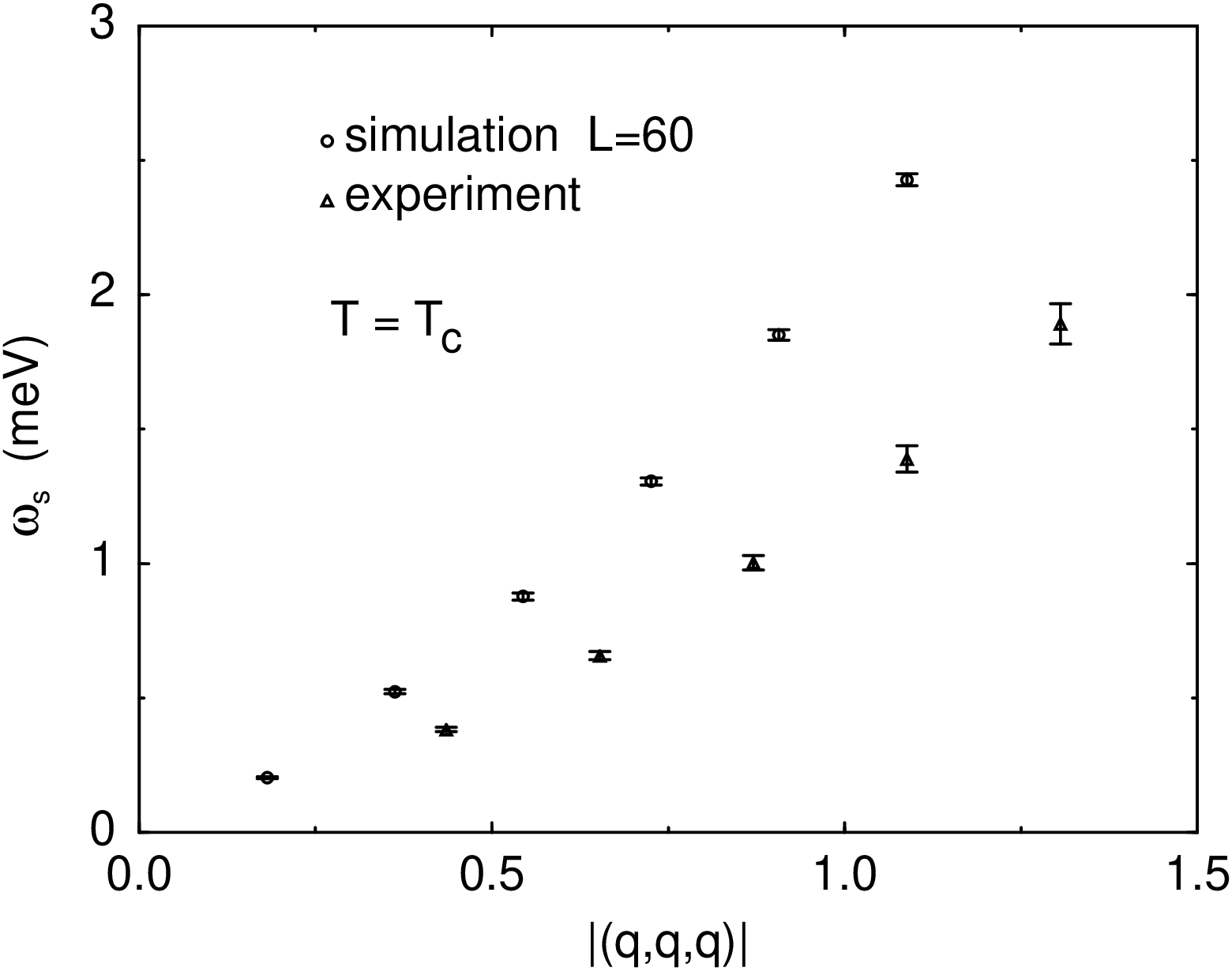}
\end{center}
\caption{}
\label{cewqTc}
\end{figure}

\begin{itemize}

\item{Fig. 1:}  Dynamic structure factor $S({\bf q},\omega)$ from our simulations
for $L=60$ at $T=0.9T_c$ and wave-vectors (a) $q=\pi/10$, (b) $\pi/6$ and (c) $\pi/3$
in the [100] direction. The symbols represent spin dynamics data and the solid line is a fit 
with the Lorentzian function given in Eq. (\ref{lorentz}). For clarity, error bars are only shown
for a few typical points, i.e. error bars for the data in the neighborhood of each of these 
points are similar. At high frequencies error bars are of the size of the fluctuations in
these data.

\item{Fig. 2:}  Dynamic structure factor $S({\bf q},\omega)$ from our simulations
for $L=60$ at $T=T_c$ and wave-vectors (a) $q=\pi/10$ and (b) $\pi/6$ in the [100] direction. 
The symbols represent spin dynamics data and the solid line is a fit with the
Lorentzian function given in Eq. (\ref{lorentz}). As in Fig. \ref{lshape09}, error bars are 
only shown for a few typical points.

\item{Fig. 3:} Spin-wave dispersion relations for $T\leq T_c$,
in the [100] direction. The symbols represent spin-wave positions extracted from Lorentzian
fits to the lineshapes from the simulations, and the solid curves are fits of the dispersion
relations at different temperatures to Eq. (\ref{wsfita0q2}).

\item{Fig. 4:} Log-log graph of the half-width of the spin-wave peak extracted from 
Lorentzian fits to the lineshapes obtained from simulations for $L=60$ and 
$T=0.9T_c$ in the [100] direction as a function of $q$.

\item{Fig. 5:} Finite-size scaling plot for $\bar\omega_m$ (with $qL=$const, 
$\delta_{\omega}L^z=$const) for the analysis with and without a resolution function. 
For the former case, the data used correspond to the converged values of $z$, for
$n=1,2$. The error bars were smaller than the symbol sizes.

\item{Fig. 6:}  Comparison of lineshapes obtained from fits to
simulation data for $L=60$ (solid line) and experiment (open circles) at $T=0.894T_c$ in 
the [111] direction: (a) $q=2\pi(0.04)$, (b) $q=2\pi(0.06)$, 
(c) $q=2\pi(0.08)$, and  (d) $q=2\pi(0.10)$. The horizontal line segment in each graph 
represents the resolution in energy (full-width at half-maximum).

\item{Fig. 7:}  Comparison of the (a) dispersion curve and 
the (b) spin-wave half-width, obtained from simulations for $L=60$ (open circle) and 
the experiment (open triangle) at $T=0.894T_c$, in the [111] direction. The simulation
data shown here correspond to values of $q$ accessible with $L=60$, without interpolation
to match the $q$ values from the experiment.

\item{Fig. 8:}  Comparison of lineshapes obtained from fits to
simulation data for $L=60$ (solid line) and experiment (open circles) at $T=T_c$ in the [111] 
direction: (a) $q=2\pi(0.04)$, (b) $q=2\pi(0.06)$, (c) $q=2\pi(0.08)$, and 
(d) $q=2\pi(0.10)$. The dot-dashed line in (c) is a fit of the experimental data to the 
functional form predicted by the RNG theory plus a Lorentzian central peak, and the RNG 
component of the fit is shown by the long-dashed line. The prediction of Mode Coupling (MC) theory 
for $q=2\pi(0.08)$ is shown by the dotted line in (c). The horizontal line segment in each graph 
represents the resolution in energy (full-width at half-maximum).

\item{Fig. 9:}  Comparison of the dispersion curve 
obtained from our simulation for $L=60$ (open circle) and the experiment (open triangle) at 
$T=T_c$, in the [111] direction. In the notation here, the first Brillouin zone edge is at 
$|(q,q,q)|\simeq 2.72$. The simulation data shown here correspond to values of $q$ accessible 
with $L=60$, without interpolation to match the $q$ values from the experiment.

\end{itemize}

\end{document}